\documentclass[9pt]{article}
\usepackage[utf8]{inputenc}
\usepackage[margin=1in]{geometry}
\usepackage{amssymb,amsmath,amsthm,amsfonts}
\usepackage{bbm}
\usepackage{natbib}
\usepackage{graphicx}
\usepackage{comment}
\usepackage{hyperref}
\usepackage{authblk}

\newcommand{\one}{\mathbbm{1}}

\graphicspath{{Images/}}
 
\title{Weighted verification tools to evaluate univariate and multivariate forecasts for high-impact weather events}
\date{}

\author[1,2]{Sam Allen}
\author[3]{Jonas Bhend}
\author[2,4]{Olivia Martius}
\author[1,2]{Johanna Ziegel}

\affil[1]{\normalsize Institute of Mathematical Statistics and Actuarial Science, University of Bern, Switzerland}
\affil[2]{\normalsize Oeschger Centre for Climate Change Research, University of Bern, Switzerland}
\affil[3]{\normalsize Swiss Federal Office of Meteorology and Climatology (MeteoSwiss)}
\affil[4]{\normalsize Institute of Geography, University of Bern, Switzerland}

\affil[ ]{\normalsize Correspondence: \url{sam.allen@stat.unibe.ch}}

\begin{document}

\maketitle

\begin{abstract}
    To mitigate the impacts associated with adverse weather conditions, meteorological services issue weather warnings to the general public. These warnings rely heavily on forecasts issued by underlying prediction systems. When deciding which prediction system(s) to utilise to construct warnings, it is important to compare systems in their ability to forecast the occurrence and severity of extreme weather events. However, evaluating forecasts for extreme events is known to be a challenging task. This is exacerbated further by the fact that high-impact weather often manifests as a result of several confounding features, a realisation that has led to considerable research on so-called compound weather events. Both univariate and multivariate methods are therefore required to evaluate forecasts for high-impact weather. In this paper, we discuss weighted verification tools, which allow particular outcomes to be emphasised during forecast evaluation. We review and compare different approaches to construct weighted scoring rules, both in a univariate and multivariate setting, and we leverage existing results on weighted scores to introduce weighted probability integral transform (PIT) histograms, allowing forecast calibration to be assessed conditionally on particular outcomes having occurred. To illustrate the practical benefit afforded by these weighted verification tools, they are employed in a case study to evaluate forecasts for extreme heat events issued by the Swiss Federal Office of Meteorology and Climatology (MeteoSwiss).
\end{abstract}

\section{Introduction}


The impacts associated with extreme weather conditions are well-documented. To mitigate these impacts, meteorological services issue weather warnings that inform the general public when hazardous conditions are expected, and outline what action should be taken to minimise the associated risks. Operational warning systems typically account not only for how likely it is that a high-impact weather event will occur, but also for other factors, such as how the public will behave in response to a warning \citep{WMO2015}. Evaluating the quality of a warning system is thus an intrinsically difficult task. However, if a warning system has access to more accurate forecasts for extreme weather events, then it has the potential to generate more useful warnings. Methods to evaluate forecasts for these high-impact events can therefore play an integral role when developing warning systems.

\bigskip


Traditionally, the evaluation of weather forecasts focuses on two aspects of forecast performance: forecast calibration and forecast accuracy. Forecast calibration considers to what extent forecasts are reliable, or trustworthy - for example, do the observed outcomes occur with the same probability with which they are predicted? This is typically assessed visually using graphical diagnostic tools, such as reliability diagrams or rank histograms \citep{Hamill2001, Jolliffe2012, Dimitriadis2021}, though statistical tests also exist to check the calibration more rigorously \citep[e.g.][]{Wilks2019, Arnold2021}. Forecast accuracy, on the other hand, is a measure of the agreement between a forecast and the corresponding observation, and is quantified using proper scoring rules. Scoring rules summarise forecast performance using a single numerical value, allowing competing forecasters to be ranked and compared objectively, and proper scoring rules encourage the forecaster to issue what they truly believe will occur \citep{GneitingRaftery2007}.

\bigskip


However, when interest is on particular outcomes, such as high-impact events, classical evaluation techniques risk raising the forecaster's dilemma; in particular,  \cite{Lerch2017} remark that ``if forecast evaluation proceeds conditionally on a catastrophic event having been observed, always predicting calamity becomes a worthwhile strategy." \cite{GneitingRanjan2011} demonstrate that a proper scoring rule is rendered improper if it is used to evaluate only the forecasts issued when particular outcomes have occurred, and \cite{Bellier2017} note that the forecaster's dilemma also applies to checks for forecast calibration. This raises questions regarding how forecasts for high-impact events should be assessed.

\bigskip


If only the occurrence of a high-impact event is of interest, then forecasts for this binary outcome can be evaluated using established verification tools: contingency table-based methods assess forecasts that are themselves binary \citep{Stephenson2008, Ferro2011}, whereas probabilistic forecasts for the event occurrence can be evaluated using reliability diagrams and appropriate scoring rules. However, relatively few methods exist to evaluate forecasts for the severity of a high-impact event. Over the past decade, the canonical approach to achieve this has been to employ weighted scoring rules, which emphasise particular outcomes during forecast evaluation whilst circumventing the forecaster's dilemma \citep{GneitingRanjan2011, Diks2011, Holzmann2017}.

\bigskip

While weighted scoring rules have mostly been applied in univariate settings, they can also be used to place more weight on certain multivariate outcomes when assessing forecast accuracy \citep{Allen2022}. The application of weighted scoring rules in a multivariate context is particularly useful when evaluating forecasts for high-impact weather events, since such events are often inherently multivariate. In particular, high-impact weather may arise not only from an extreme event, but also from the interaction of several more moderate events; this has been the catalyst for numerous recent studies on so-called compound weather events \citep[see][for a review]{Zscheischler2020}. 

\bigskip


Various approaches to construct weighted scoring rules have been proposed. In this paper, we discuss and compare these approaches, and provide guidance regarding which should be employed in different circumstances, both in a univariate and multivariate setting. While weighted scoring rules provide a measure of forecast accuracy when predicting extreme events, we demonstrate that the theory underlying these weighted scores can readily be applied to checks for forecast calibration. Moreover, we introduce a novel diagnostic tool that can assess the calibration of probabilistic forecasts conditionally on particular outcomes having occurred. 

\bigskip


To demonstrate how they can be applied in practice, these weighted verification tools are applied to forecasts of extreme heat. Extreme heat provides a salient example of a compound weather event: while short and intense periods of extreme heat can have serious implications for human health (among other things), persistent hot periods further strain the human body by inhibiting its ability to recover \citep{Basagana2011}. Most major weather services therefore issue warnings to the public when persistently high temperatures are expected, and a greater understanding of how well such events can be predicted would allow weather services to further refine their heat warning systems.

\bigskip


The remainder of the paper is structured as follows. In the following section, we review the general framework for forecast evaluation and introduce relevant weighted verification tools when evaluating forecasts for high-impact weather events. These are then applied to forecasts of extreme heat events in Section \ref{section:casestudy}. In particular, operational forecasts issued by the Swiss Federal Office of Meteorology and Climatology (MeteoSwiss) are compared to forecasts obtained from a statistical post-processing model, also introduced in Section \ref{section:casestudy}, allowing us to analyse the effect of post-processing when forecasting extreme heat. The conclusions drawn from this case study are discussed in Section \ref{section:conclusion}.

\bigskip

\section{Forecast verification}
\label{section:verification}

\subsection{Forecast accuracy}


Suppose we are interested in forecasting a random variable $ Y $ that takes values in a set $ \Omega $, and that our forecasts are probability distributions on $ \Omega $. Let $ \mathcal{F} $ denote a set of such distributions. A scoring rule $ S $ is a function that takes a forecast $ F \in \mathcal{F} $ and an observation $ y \in \Omega $ as inputs, and outputs a numerical value, or score, that quantifies the forecast accuracy. All scoring rules considered herein are negatively oriented, so that a more accurate forecast receives a lower score. A scoring rule $ S $ is called proper with respect to $ \mathcal{F} $ if $ \mathbbm{E}_{G}[S(G, Y)] \leq \mathbbm{E}_{G}[S(F, Y)] $ for all $ F, G \in \mathcal{F} $, where $ \mathbbm{E}_{G} $ denotes the expectation over $ G $, and strictly proper with respect to $ \mathcal{F} $ if this holds with equality if and only if $ F = G $. That is, if the observations are believed to arise according to a certain distribution, then the expected score is optimised when this distribution is issued as the forecast. We assume throughout that the expectations are finite where necessary.

\bigskip

Proper scoring rules exist to assess forecasts for a range of different outcomes \citep{GneitingRaftery2007}. When considering high-impact events, it is common to reduce the problem to a binary forecasting task, whereby we are only interested in predicting whether or not the event of interest will occur. Although forecasts for such events could themselves be binary, it is more natural to issue forecasts that are probabilistic, thereby quantifying the uncertainty inherent in the prediction. One of the most popular scoring rules to evaluate such forecasts is the Brier score \citep{Brier1950}. Consider the case where the outcome is univariate and real-valued, i.e.\ $ \Omega = \mathbbm{R} $, and the forecast $ F $ is a cumulative distribution function over the real line. A high-impact event might then be defined as an instance where the outcome exceeds a certain threshold $ t $, in which case the Brier score is defined as
\begin{equation}\label{eq:BS}
    \mathrm{BS}(F, y; t) = (F(t) - \one\{y \leq t\})^{2},
\end{equation}
where $ \one $ denotes the indicator function.

\bigskip

Of course, in considering only the occurrence of a high-impact event, we cannot assess how well forecasts predict the event's severity. While the Brier score above evaluates the forecast at a particular threshold, the entire forecast distribution can be evaluated by integrating the Brier score over all possible thresholds. In doing so, we obtain the continuous ranked probability score (CRPS), the most commonly used scoring rule to evaluate probabilistic forecasts. If our forecast distribution $ F $ has a finite mean, the CRPS can be expressed as
\begin{equation}
\begin{split}\label{eq:crps}
    \mathrm{CRPS}(F, y) & = \int_{-\infty}^{\infty} (F(z) - \one\{y \leq z\})^{2} \mathrm{d}z \\
    & = \mathbbm{E}_{F}|X - y| - \frac{1}{2}\mathbbm{E}_{F}|X - X^{\prime}|,
\end{split}
\end{equation}
where $ X $ and $ X^{\prime} $ are independent random variables that follow the distribution $ F $ \citep{Matheson1976, GneitingRaftery2007}. 

\bigskip

The CRPS is regularly employed in climate-related studies, in part because it can readily be applied to ensemble forecasts by replacing the expectations in the second expression of Equation \ref{eq:crps} with sample means over the ensemble members. The CRPS assesses the forecast distribution over the set of all possible outcomes, providing a measure of overall forecast performance, rather than evaluating forecasts made for high-impact events. Nonetheless, several extensions of the CRPS have been proposed that can emphasise particular outcomes whilst assessing forecast accuracy. 

\bigskip

\subsubsection*{Weighted scoring rules}

In order to emphasise high-impact events during forecast evaluation, a seemingly intuitive approach would be to only evaluate the forecasts issued when such an event occurs; more generally, to assign a higher weight to outcomes corresponding to higher impacts. However, \cite{GneitingRanjan2011} demonstrate that if proper scoring rules are weighted by a function that depends on the observed outcome, then the score is generally rendered improper. For example, if evaluation is restricted to instances where high-impact events occur, then the forecaster is encouraged to always predict that such an event will occur, even though such a forecast is uninformative in practice \citep{Lerch2017}.

\bigskip

Instead, weighted scoring rules have been introduced to target particular outcomes during forecast evaluation in a more theoretically desirable way. Weighted scoring rules incorporate a weight function into conventional scoring rules, but do such in such a way that the resulting score remains proper. The weight function can then be chosen to emphasise particular outcomes of interest. In the following, a weight function is a function $ w $ such that $ w(z) \geq 0 $ for all possible outcomes $ z $.  \cite{GneitingRanjan2011} list weight functions that could be used to emphasise certain real-valued outcomes, and these are given in Table \ref{tab:mv_weights}.

\bigskip

Having chosen a suitable weight function for the problem at hand, several approaches have been proposed to incorporate this weight into existing scoring rules. \cite{Allen2022} list three possible methods to emphasise particular outcomes when evaluating forecasts using the CRPS. Firstly, the threshold-weighted CPRS (twCRPS) introduced by \cite{Matheson1976} and \cite{GneitingRanjan2011} is defined as
\begin{equation}
\begin{split}\label{eq:twcrps_integral}
    \mathrm{twCRPS}(F, y; w) & = \int_{-\infty}^{\infty} (F(z) - \one\{y \leq z\})^{2} w(z) \mathrm{d}z, \\
    & = \mathbbm{E}_{F}|v(X) - v(y)| - \frac{1}{2}\mathbbm{E}_{F}|v(X) - v(X^{\prime})|,
\end{split}
\end{equation}
where $ X, X^{\prime} \sim F $ are independent, and $ v $ is any function such that $ v(z) - v(z^{\prime}) = \int_{z^{\prime}}^{z} w(x) \mathrm{d}x $ and $ \mathbbm{E}_{F}|v(X)| < \infty$. Secondly, \cite{Holzmann2017} proposed the outcome-weighted CRPS (owCRPS):
\begin{equation}\label{eq:owcrps}
\begin{split}
    \mathrm{owCRPS}(F, y; w) & = w(y)\int_{-\infty}^{\infty} (F_{w}(z) - \one\{y \leq z\})^{2} \mathrm{d} z, \\
    & = w(y) \mathrm{CRPS}(F_{w}, y),
\end{split}
\end{equation}
where 
\begin{equation}\label{eq:weightedF}
    F_{w}(x) = \frac{\mathbb{E}_{F}\left[ \one\{ X \leq x\} w(X) \right] }{\mathbb{E}_{F}[w(X)]},
\end{equation}
with $ X \sim F $. Lastly, \cite{Allen2022} introduced the vertically re-scaled CRPS (vrCRPS):
\begin{equation}\label{eq:vrcrps}
\begin{split}
    \mathrm{vrCRPS}(F, y; w, x_{0}) = & \mathbb{E}_{F}[|X - y|w(X)w(y)] - \frac{1}{2}\mathbb{E}_{F}[|X - X^{\prime}|w(X)w(X^{\prime})] \\
    & + (\mathbb{E}_{F}[ |X - x_{0}|w(X) ] - |y - x_{0}|w(y) ) (\mathbb{E}_{F}[w(X)] - w(y)),
\end{split}
\end{equation}
where $ X, X^{\prime} \sim F $ are independent, and $ x_{0} $ is an arbitrary real value. \cite{GneitingRanjan2011} additionally introduced a quantile-weighted version of the CRPS, though this emphasises particular regions of the forecast distribution rather than particular outcomes, and is thus not considered here. 

\bigskip

In all cases, the unweighted CRPS is recovered when the weight function is constant and equal to one. Of the above three approaches to weight the CRPS, the twCRPS is the most well-known, and has been applied in several studies to evaluate forecasts for extreme weather events \citep[e.g.][]{Lerch2013, Allen2021}. The outcome-weighted CRPS, on the other hand, has been used to assess economic forecasts, but is relatively unknown within the field of weather and climate forecasting. An obvious question then is how these weighted scores differ from one another, and which (if any) should be preferred when evaluating forecasts for high-impact weather events? In this section, we seek to answer this question by providing a detailed comparison of the different approaches.
 
\bigskip

Firstly, consider how these weighted scores differ. As discussed, the CRPS is defined as an integral of the Brier score when predicting whether the observation will exceed a certain threshold, and the twCRPS simply assigns different weights to different thresholds in the integration. Note that the weight in the twCRPS depends on the variable of integration, rather than the observation. The second expression in Equation \ref{eq:twcrps_integral} demonstrates that the twCRPS can additionally be interpreted as the CRPS after having transformed the forecasts and observations, with the transformation $ v $ - which \cite{Allen2022} call the chaining function - governed by the choice of weight function. 

\bigskip

In contrast to the twCRPS, the owCRPS employs a weight that depends on the outcome. \cite{GneitingRanjan2011} demonstrate that if the CRPS is weighted by a function that depends on the observed outcome, then the expectation of this weighted score, i.e.\ $ \mathbbm{E}_{G}[w(Y) \mathrm{CRPS}(F, Y)] $ with $ Y \sim G $, is minimised by issuing $ G_{w} $ as the forecast, rather than $ G $, where $ G_{w} $ is defined analogously to $ F_{w} $ in Equation \ref{eq:weightedF}. This weighted scoring rule is therefore generally improper. To circumvent this, \cite{Holzmann2017} suggest evaluating the forecasts via their weighted representation, providing an arguably more direct way of circumventing the forecaster's dilemma than the twCRPS.

\bigskip

\begin{figure}
    \centering
    \includegraphics[width=0.3\textwidth]{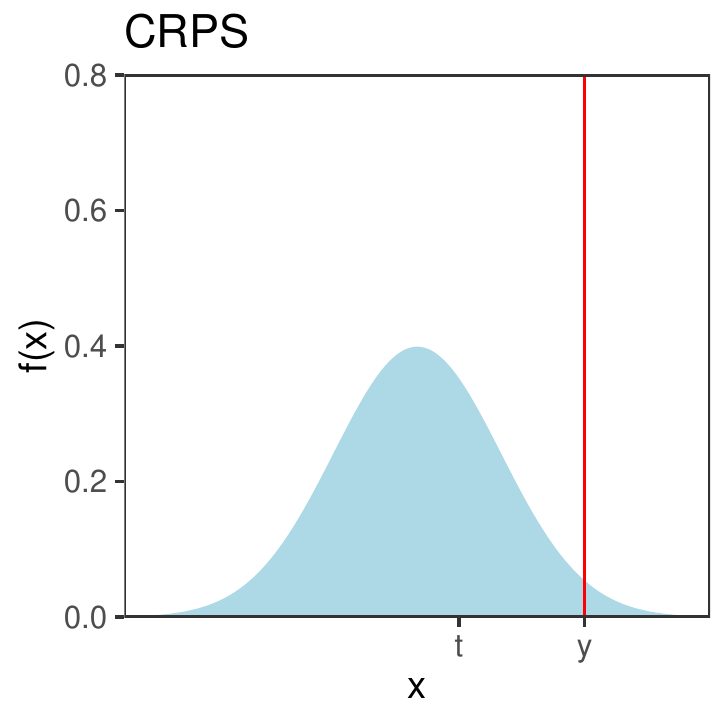}
    \includegraphics[width=0.3\textwidth]{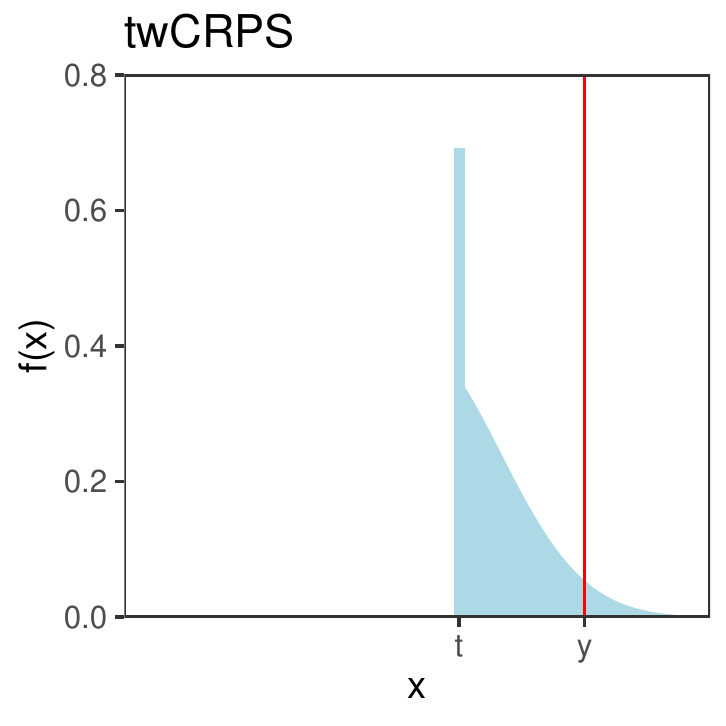}
    \includegraphics[width=0.3\textwidth]{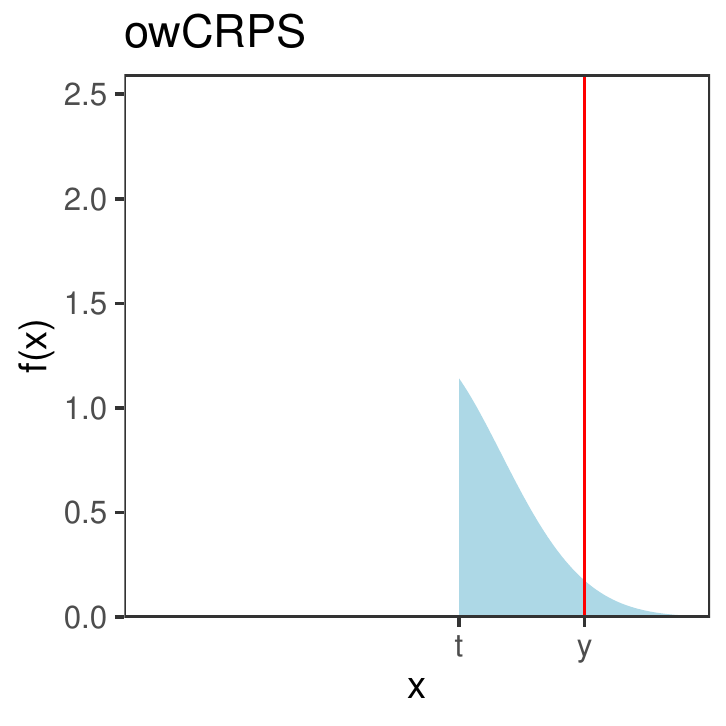}
    \caption{Probability density functions used when calculating the CRPS, twCRPS, and owCRPS, with a weight function equal to $ w(z) = \one\{z > t\} $. Note the contrasting scale for the owCRPS.}
    \label{fig:dist_demo}
\end{figure}

Both the twCRPS and the owCRPS transform the forecasts and observations prior to implementing the unweighted CRPS, with the two approaches differing in the transformation they employ. Consider the common case where the weight function restricts attention to values above some threshold of interest $ t $, i.e.\ $ w(z) = \one\{z > t\} $. Figure \ref{fig:dist_demo} illustrates the difference between these two transformations for such a weight function. While the CRPS measures the distance between the observation and the entire forecast distribution, the twCRPS reassigns all probability assigned to values lower than the threshold to the threshold itself. This results in a left-censored distribution, with a point mass at the threshold of interest. In doing so, the score only depends on how the forecast behaves above the threshold. The owCRPS, on the other hand, truncates the distribution at the threshold, thereby evaluating the conditional distribution given that the threshold has been exceeded. This relies on the observation exceeding the threshold, and the owCRPS is zero whenever this is not the case.

\bigskip

In considering this conditional distribution, the owCRPS is only sensitive to the shape of the forecast distribution above the threshold, and not to the forecast probability that the threshold will be exceeded; that is, the score cannot distinguish between two forecasts that have the same conditional distribution. It therefore only assesses the predicted severity of a high-impact event, whereas the twCRPS additionally accounts for the probability with which the event is predicted to occur. \cite{Holzmann2017} suggest complementing the owCRPS by adding the score to a scoring rule for binary events, such as the Brier score, which can independently evaluate the probability forecasts. For example,
\begin{equation}\label{eq:owcrps+bs}
\begin{split}
    \mathrm{owCRPS_{(BS)}}(F, y; w) & = \mathrm{owCRPS}(F, y; w) + \one\{ y \leq t \} (F(t) - \one\{y \leq t\})^{2} \\
    & = \one\{ y > t \} \mathrm{CRPS}(F_{w}, y) + \one\{ y \leq t \} \mathrm{BS}(F, y; t).
\end{split}
\end{equation}
A similar extension of the owCRPS is also possible when alternative weight functions are considered \citep{Holzmann2017}.

\bigskip

However, even when complemented with such a score, the owCRPS does not consider the shape of the forecast distribution when the outcome does not exceed the threshold: if two forecasts assign the same probability to the exceedance of the threshold, then they will receive the same score, even if one predicts more severe events with a higher probability. Instead, this binary score could be replaced with a score that also accounts for the distance between the probability distribution and the threshold, penalising forecast distributions that assign higher probabilities to values much larger than the threshold. It turns out that this is in essence what the twCRPS does. In fact, if $ w(z) = \one\{z > t\} $, it is straightforward to rewrite the twCRPS in terms of the owCRPS:
\begin{equation}
\begin{split}
    \mathrm{twCRPS}(F, y; w) = &  (1 - F(t))^{2} \mathrm{owCRPS}(F, y; w) \\
    & + \one\{ y > t\} \left[F(t)^{2} (y - t) + 2F(t)\int_{t}^{y} F(x) - F(t) \mathrm{d}x \right] \\
    & + \one \{ y \leq t \} \int_{t}^{\infty} (F(x) - 1)^{2} \mathrm{d}x.    
\end{split}
\end{equation}
For this weight function, the twCRPS thus differs from the owCRPS in two main respects. Firstly, the twCRPS depends on the forecast even if the outcome does not exceed the threshold of interest, whereas the owCRPS is always equal to zero. While complemented versions of the owCRPS (e.g. Equation \ref{eq:owcrps+bs}) address this, the twCRPS accounts not only for the probability that the threshold will be exceeded, but also the distance from the forecast distribution to this threshold. Secondly, when the threshold is exceeded by the outcome, the twCRPS is additionally comprised of two terms not present in the owCRPS, both of which penalise forecasts that issue a high probability that the outcome will not exceed the threshold. 

\bigskip

The vrCRPS differs from the other two scores in that it does not transform the forecasts and observations, but rather weights the distance between them. In doing so, the vrCRPS depends not only on a weight function, but also on an additional parameter $ x_{0} $. Although this could be construed as a practical disadvantage of the score, \cite{Allen2022} note that when $ w(z) = \one\{z > t\} $, a canonical choice for this parameter is $ x_{0} = t $, in which case the vrCRPS is in fact equivalent to the twCRPS. When there is no canonical choice for $ x_{0} $, it can arbitrarily be set equal to zero.

\bigskip

\begin{figure}
    \centering
    \includegraphics[width=0.5\textwidth]{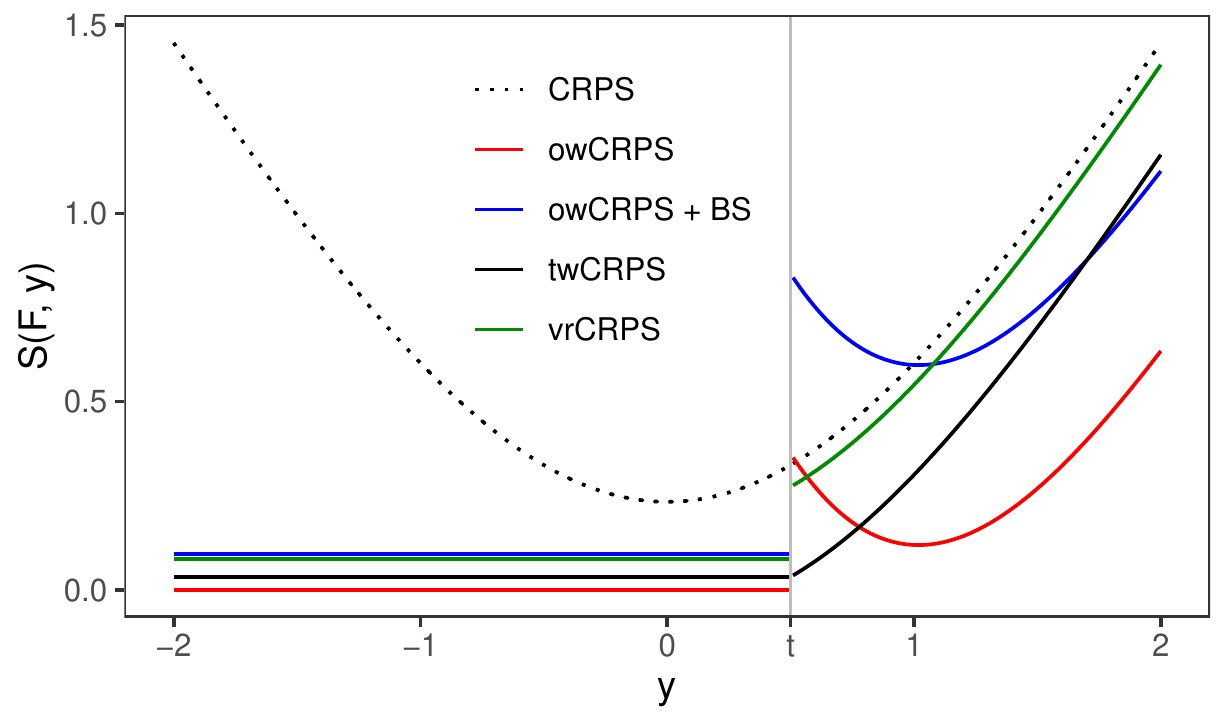}
    \caption{CRPS and weighted versions of the CRPS for a standard normal distribution as a function of the observation $ y $. The weight function $ w(z) = \one\{ z > t\} $ is used within the weighted scores, and a vertical grey line is shown at $t$.}
    \label{fig:score_demo}
\end{figure}

For this indicator-based weight function, a simple illustration of how the weighted scores behave is displayed in Figure \ref{fig:score_demo}. The forecast in this case is a standard normal distribution, and the scores are shown as a function of the observation. The CRPS clearly increases as the observed value moves away from the forecast mean, while all weighted scores are constant when the observation falls below the threshold of interest. The twCRPS and vrCRPS are proportional to the CRPS above this threshold, and the twCRPS has the desirable property that it is continuous: there is a jump in all other scores at the threshold, meaning a small difference in the observation can lead to a large change in the score. For the vrCRPS, the magnitude of this difference is controlled by $ x_{0} $. When $ x_{0} = t $, the vrCRPS is also continuous, since the score is equivalent to the twCRPS (for comparison, the vrCRPS shown in Figure \ref{fig:score_demo} employs $ x_{0} = 0 $). 
\bigskip

The twCRPS and vrCRPS additionally have the benefit that they can readily be applied to ensemble forecasts, or forecasts in the form of Monte-Carlo samples; as with the CRPS, this can be achieved by replacing the expectations in their definitions with sample means over the ensemble members \cite[see also][]{Allen2022}. Note that for the twCRPS, the chaining function $ v $ is typically straightforward to calculate for the weights frequently employed in practice. The owCRPS, on the other hand, relies on the weighted forecast distribution $ F_{w} $ being well-defined (i.e.\ $ \mathbbm{E}_{F}[w(X)] > 0 $), which is often not the case if the weight function targets extreme events and the forecast is an ensemble. In this case, it may be necessary to smooth the ensemble to form a continuous forecast distribution prior to assessing the forecasts. Although the use of strictly positive weight functions would ensure $ \mathbbm{E}_{F}[w(X)] > 0 $ in theory, this can still lead to numerical complications in practice. Hence, when interest is on high-impact weather events, we recommend evaluating forecast accuracy using the twCRPS or vrCRPS. 

\bigskip

\subsubsection*{Multivariate weighted scoring rules}

Since high-impact weather often arises as a combination of weather events across multiple dimensions, forecasts for such events should be assessed using both univariate and multivariate techniques. Suppose now that $ \Omega = \mathbbm{R}^{d} $, for $ d > 1$, and that the forecast $ F $ is a probability distribution on $ \mathbbm{R}^{d} $. Two of the most popular scoring rules to assess such forecasts are the energy score \citep[ES;][]{GneitingRaftery2007} and the variogram score \citep[VS;][]{ScheuererHamill2015}. The energy score is defined as
\begin{equation}\label{eq:ES}
    \mathrm{ES}(F, y) = \mathbbm{E}_{F}||X - y|| - \frac{1}{2}\mathbbm{E}_{F}||X - X^{\prime}||,
\end{equation}
where $ || \cdot || $ is the Euclidean distance in $ \mathbbm{R}^{d} $, and $ X, X^{\prime} \sim F $ are independent \citep{GneitingRaftery2007}. The energy score provides a natural generalisation of the CRPS to higher dimensions, and is commonly employed in practice since it can readily be applied to ensemble forecasts.

\bigskip

However, several studies have demonstrated that the energy score may fail to be discriminative when comparing forecasts with different dependence structures \citep[e.g.][]{Pinson2013}. \cite{ScheuererHamill2015} introduce the variogram score as an alternative scoring rule, which exploits the variogram commonly used in spatial statistics in order to directly target the forecast's multivariate dependence structure. The variogram score of order $ p > 0 $ is defined as
\begin{equation}\label{eq:VS}
    \mathrm{VS}_{p}(F, y) = \sum_{i=1}^{d}\sum_{j=1}^{d} h_{i, j} (\mathbb{E}_{F}|X_{i} - X_{j}|^{p} - |y_{i} - y_{j}|^{p})^{2},
\end{equation}
where $ y = (y_{1}, \dots, y_{d}) \in \mathbbm{R}^{d} $, $ X = (X_{1}, \dots, X_{d}) \sim F $, and $ h_{i, j} \in [0, 1] $ are non-negative scaling parameters. In the following, $ p $ is chosen to be one half, as recommended by \cite{ScheuererHamill2015}, and the weights $ h_{i, j} $ are all set to one.

\bigskip

As in the univariate case, weighted versions of these scores exist that allow particular outcomes to be targeted during forecast evaluation. In this case, the weight functions should be defined on $ \mathbbm{R}^{d} $ rather than the real line. \cite{GneitingRanjan2011} propose several univariate weight functions based on Gaussian density and distribution functions, and weights to emphasise certain regions of the multivariate outcome space can be defined analogously in terms of multivariate Gaussian density and distribution functions. Some examples of such multivariate extensions are listed in Table \ref{tab:mv_weights}. Of course, alternative weight functions could also be applied, and the most appropriate weight will depend on what information is to be extracted from the forecasts during evaluation.

\bigskip

\begin{table}
\centering
    \begin{tabular}{| c | c | c |}
    \hline
        Values of interest & Univariate weight & Multivariate weight \\
        \hline
        All values & $ w(x) = 1 $ & $ w(x) = 1 $ \\
        Central values & $ w(x) = \phi_{\mu, \sigma}(x) $ & $ w({x}) = \boldsymbol{\phi_{\mu, \Sigma}}(x) $ \\
        Tail values & $ w(x) = 1 - \phi_{\mu, \sigma}(x)/\phi_{\mu, \sigma}(\mu) $ & $ w(x) = 1 - \boldsymbol{\phi_{\mu, \Sigma}}(x)/\boldsymbol{\phi_{\mu, \Sigma}}(\boldsymbol{\mu}) $ \\
        Right tail/Upper right quadrant & $ w(x) = \Phi_{\mu, \sigma}(x) $ & $ w(x) = \boldsymbol{\Phi_{\mu, \Sigma}}(x) $ \\ 
        Left tail/Lower left quadrant & $ w(x) = 1 - \Phi_{\mu, \sigma}(x) $ & $ w(x) = 1 - \boldsymbol{\Phi_{\mu, \Sigma}}(x) $ \\ 
        \hline
    \end{tabular}
    \caption{Possible weight functions for univariate and multivariate weighted scores. Here, $ \phi_{\mu, \sigma} $ and $ \Phi_{\mu, \sigma} $ denote the density and distribution functions, respectively, of the Gaussian distribution with mean $ \mu $ and standard deviation $ \sigma $, while $ \boldsymbol{\phi_{\mu, \Sigma}} $ and $ \boldsymbol{\Phi_{\mu, \Sigma}} $ denote the density and distribution functions, respectively, of the multivariate Gaussian distribution with mean vector $ \boldsymbol{\mu} $ and covariance matrix $ \boldsymbol{\Sigma} $.}
    \label{tab:mv_weights}
\end{table}

The three approaches to generate weighted versions of the CRPS can also be applied to other scoring rules. It is possible to construct an outcome-weighted version of any proper score \citep{Holzmann2017}, while threshold-weighting and vertically re-scaling are applicable to the very general class of kernel scores \citep{GneitingRaftery2007, Allen2022}. Since the energy score and variogram score both belong to the class of kernel scores, it is possible to introduce threshold-weighted, outcome-weighted, and vertically re-scaled versions of these multivariate scores, which can emphasise particular multivariate outcomes when evaluating forecast accuracy \citep{Allen2022}. For example, threshold-weighted energy and variogram scores can be defined as follows:
\begin{equation}\label{equation:twES}
    \mathrm{twES}(F, y; v) = \mathbbm{E}_{F}||v(X) - v(X^{\prime})|| - \frac{1}{2}\mathbbm{E}_{F}||v(X) - v(X^{\prime})||,
\end{equation}
\begin{equation}\label{eq:twVS}
    \mathrm{twVS}_{p}(F, y; v) = \sum_{i=1}^{d}\sum_{j=1}^{d} h_{i, j} (\mathbb{E}_{F}|v(X)_{i} - v(X)_{j}|^{p} - |v(y)_{i} - v(y)_{j}|^{p})^{2},
\end{equation}
where $ X, X^{\prime} \sim F $ are independent, and $ v: \mathbbm{R}^{d} \to \mathbbm{R}^{d} $. As with the twCRPS, these scores involve a transformation of the forecasts and observations prior to calculating the unweighted scores. Outcome-weighted and vertically re-scaled versions of these scores can similarly be introduced \citep[see][for details]{Allen2022}.

\bigskip

We can again consider how these weighted scores differ. Firstly note that, although the energy score and variogram score are arguably the most popular scoring rules to evaluate multivariate weather forecasts, other multivariate scoring rules exist, such as the logarithmic score and the Dawid-Sebastiani score \citep{DawidSebastiani1999}. While it is possible to construct outcome-weighted versions of these scores, these scores do not fit into the kernel score framework, and hence threshold-weighted and vertically re-scaled versions of these scores cannot readily be defined. This approach of outcome-weighting is therefore more general than threshold-weighting and vertically re-scaling. However, the outcome-weighted multivariate scores again rely on $ \mathbbm{E}_{F}[w(X)] $ being non-zero, and since multivariate weather forecasts almost exclusively take the form of ensembles, implementing outcome-weighted scores to evaluate forecasts for high-impact weather events becomes yet more challenging in a multivariate setting.

\bigskip 

The threshold-weighted multivariate scores are defined in terms of a chaining function $ v $ that is used to transform the forecasts and observations. However, in contrast to the univariate case, there is no general framework with which to obtain a chaining function from a weight function on $ \mathbbm{R}^{d} $. \cite{Allen2022} show that if the weight function is always equal to either zero or one, then a canonical choice for the chaining function is 
\begin{align}\label{equation:chaining}
v(z) = 
\begin{cases}
  z  & \text{ if } w(z) = 1, \\
  z_{0} & \text{ if } w(z) = 0,
 \end{cases}
\end{align}
where $ z_{0} $ is an arbitrary point in $ \mathbbm{R}^{d} $. With such a weight function, the score will depend only on how the forecast distribution behaves at points $ z $ for which $ w(z) = 1 $. However, for more general weight functions, there is no obvious and general framework to construct a chaining function from a weight, and choosing a chaining function to emphasise the events of interest is somewhat less intuitive than selecting an appropriate weight function.

\bigskip

Conversely, the vertically re-scaled energy and variogram scores depend directly on a multivariate weight function. As a result, they can readily be applied with arbitrarily complex weight functions, without having to additionally define a relevant chaining function. This is a practical advantage of these weighted scores. Moreover, as in the univariate case, the vertically re-scaled scores are the same as the threshold-weighted scores for particular choices of the weight and chaining functions, and both classes of weighted scores can easily be applied to multivariate ensemble forecasts. Hence, due to the ease with which they can be implemented in practice, we generally recommend using vertically re-scaled multivariate scores to emphasise particular outcomes in multiple dimensions, though threshold-weighted scores are also appealing if a canonical choice of the chaining function exists.

\bigskip

\subsection{Forecast calibration}
\label{section:calibration}

Although proper scoring rules allow competing prediction systems to be ranked and compared objectively, they cannot be used to determine whether a prediction system is trustworthy, in the sense that the observed outcomes are statistically consistent with the forecasts that were issued. If the forecasts do align with the observations, then the prediction system is said to reliable, or calibrated. 

\bigskip

When the outcomes are univariate and real-valued, the most popular tool to assess forecast calibration is the rank or probability integral transform (PIT) histogram \citep{Dawid1984, Gneiting2007}. PIT histograms rely on the result that if the outcome $ Y $ is a continuous random variable with cumulative distribution function $ F $, then $ F(Y) $ will follow the standard uniform distribution; a simple extension of this result exists for when $ Y $ is not continuous. Hence, to check for calibration, we can evaluate each forecast distribution function at the observed outcome, $ F(y) $, and display these values in a histogram. If the observations are indeed draws from the corresponding forecast distributions, then the resulting histogram should be uniform. If the histogram is not uniform, then there is evidence to suggest the forecasts are miscalibrated, and the behaviour of the deviations can be used to diagnose the nature of the forecast errors \citep{Hamill2001}. 

\bigskip
 
Although forecast calibration in this setting could also be visualised using other techniques, one reason why PIT histograms are so commonly applied in practice is because there exists a discrete analogue when forecasts are in the form of an ensemble. So-called rank histograms display the relative frequency of the rank of the observation when pooled among the corresponding ensemble members \citep[e.g.][]{Hamill1997}. If the prediction system is calibrated, then the observation should be equally likely to assume any rank on average, resulting in a uniform rank histogram. As with PIT histograms, the forecast miscalibration can be quantified by measuring the deviation between the observed histogram and a uniform histogram, and statistical tests for forecast calibration can then be derived by assessing whether or not this deviation is significantly large \citep{DelleMonache2006, Wilks2019, Arnold2021}. 

\bigskip

\subsubsection*{Conditional PIT histograms}

The forecaster's dilemma also applies to diagnostic checks for calibration: if a rank or PIT histogram is constructed from only the forecasts issued when a high-impact event occurs, then the resulting histogram of a calibrated prediction system will in general not be uniform. For this reason, \cite{Bellier2017} ``strongly advise against observation-based stratification when constructing rank histograms." While weighted scoring rules have been proposed to emphasise particular outcomes when calculating forecast accuracy, no similar extensions have been introduced when assessing forecast calibration. In this section, we leverage the previous discussion on weighted scoring rules to introduce conditional PIT (cPIT) histograms, which can be used to check the calibration of probabilistic forecasts conditionally on certain outcomes having occurred.

\bigskip

When interest is on particular real-valued outcomes, the outcome-weighted CRPS evaluates the conditional forecast distribution given that these outcomes have occurred. This conditional distribution can similarly be used within PIT histograms in order to assess forecast calibration for high-impact events. For example, let the outcome $ Y $ be a continuous (real-valued) random variable with distribution function $ F $, and let $ Y_{>t} $ denote the conditional outcome variable given that the outcome exceeds a threshold $ t $; that is, $ Y_{>t} $ follows the distribution $ G $, where $ G(x) = [F(x) - F(t)]/[1 - F(t)] $ for $ x > t $, and $ G(x) = 0 $ otherwise. The probability integral transform $ G(Y_{>t}) $ then follows a standard uniform distribution. Hence, to evaluate the calibration of forecasts conditionally on the threshold being exceeded, we can  calculate the conditional PIT values $ G(y) = [F(y) - F(t)]/[1 - F(t)] $ for all observations $ y $ that exceed $ t $, and display these values in a histogram. If the conditional distribution of $ Y $ is indeed the conditional distribution predicted by the forecasts, then the resulting histogram should be uniform, in which case the prediction system is said to be conditionally calibrated. 

\bigskip

\begin{figure}
    \centering
    \includegraphics[width=0.3\textwidth]{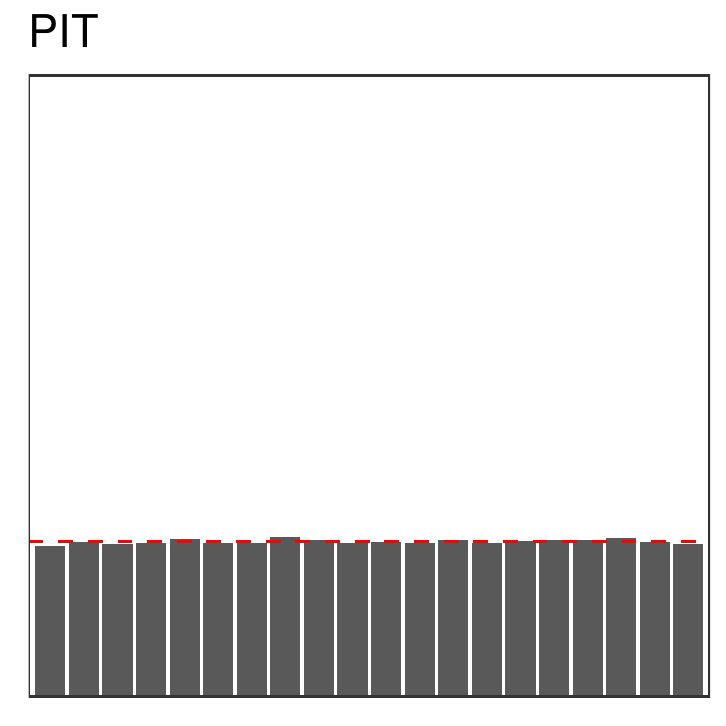}
    \includegraphics[width=0.3\textwidth]{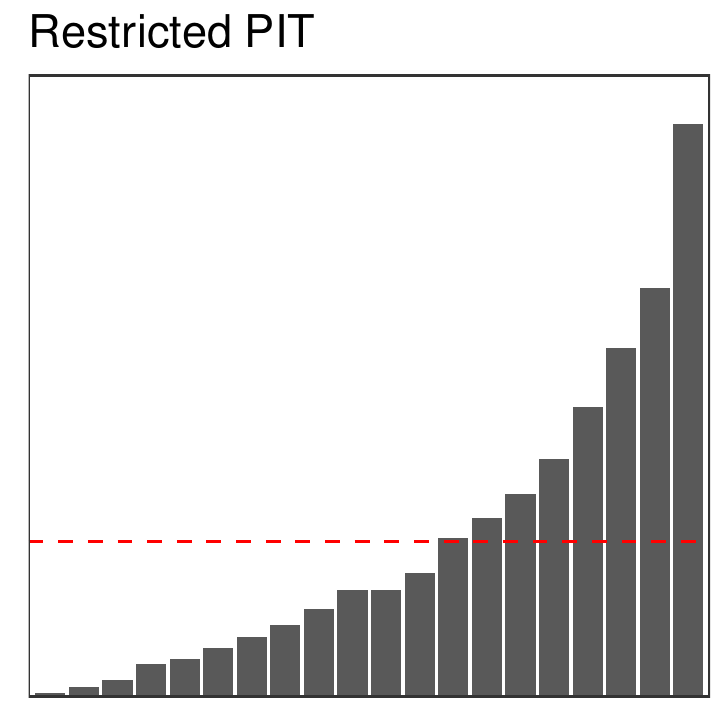}
    \includegraphics[width=0.3\textwidth]{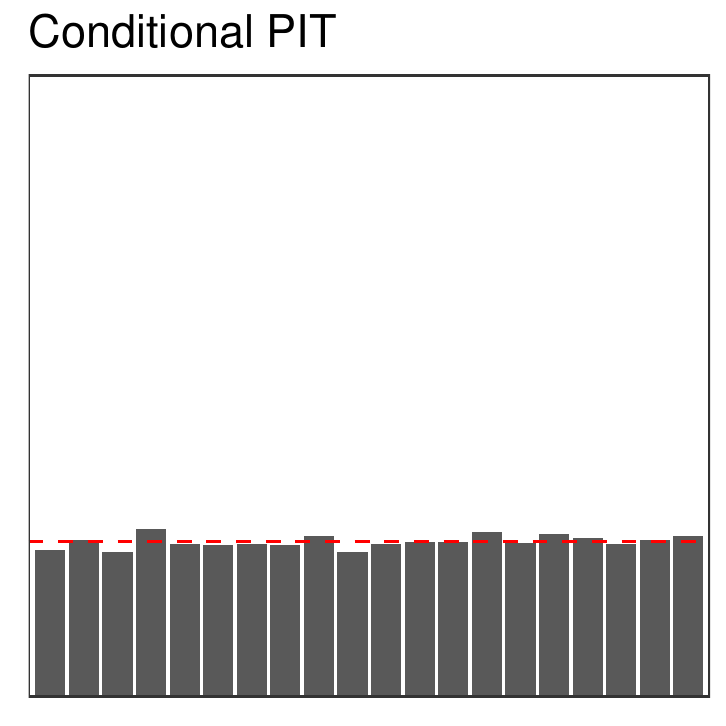}
    \caption{PIT and conditional PIT histograms for an ideal forecaster, along with the PIT histogram constructed from only the observations that exceed a certain threshold. The histograms are comprised of 100,000 observations from a $ \mathcal{N}(\mu, \sigma^{2}) $ distribution, where $ \mu \sim \mathcal{N}(0, 1 - \sigma^{2}) $ and $ \sigma^{2} = 1/3 $. A threshold of $ t = 1$ is used 
    within the weighted histograms.}
    \label{fig:PIT_demo}
\end{figure}

These cPIT histograms are not equivalent to focusing on the bins on the right-hand side of the standard PIT histogram, since an extreme observation could correspond to a low PIT value $ F(y) $ if the forecast predicts more extreme events to occur with a high probability. To illustrate this, Figure \ref{fig:PIT_demo} displays PIT and cPIT histograms for a perfect or ideal prediction system, as well as a histogram comprised of the PIT values that correspond to observations above a threshold of interest (labelled a restricted PIT histogram). The prediction system is calibrated, resulting in a uniform PIT histogram, but when interest is restricted to observations that exceed the threshold, the histogram becomes considerably skewed. The cPIT histogram, on the other hand, remains uniform, suggesting the forecasts are conditionally calibrated.

\bigskip

In theory, if the prediction system is calibrated, then it will additionally be conditionally calibrated, irrespective of the outcomes considered in the conditional PIT histograms. However, in practice, a forecast that appears calibrated may be significantly miscalibrated when more focus is put on particular outcomes. An example of this is presented in Section \ref{section:casestudy}. Conversely, a forecast that is miscalibrated overall, leading to a non-uniform PIT histogram, may still be calibrated conditionally on the occurrence of a high-impact event. 

\bigskip

\begin{figure}
    \centering
    \raisebox{0.9cm}{\includegraphics[width=0.29\textwidth]{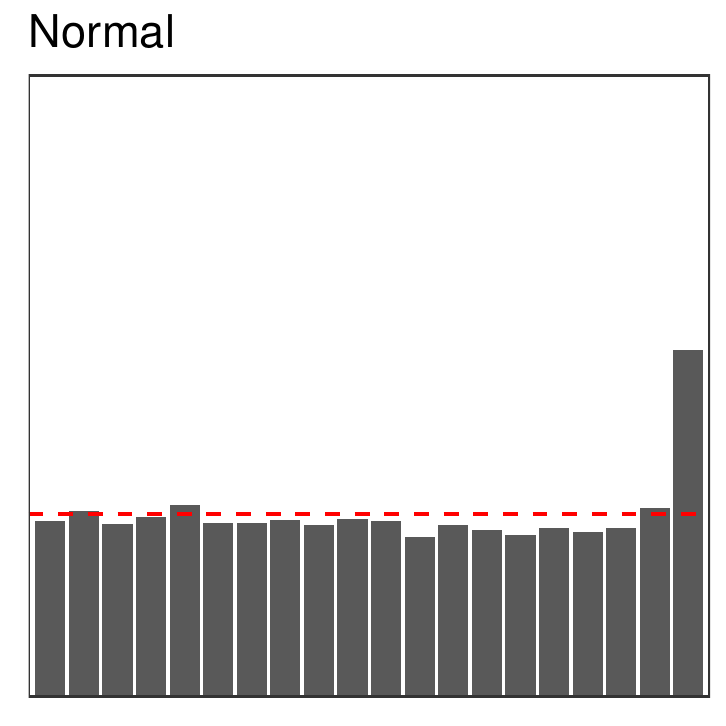}}
    \includegraphics[width=0.38\textwidth]{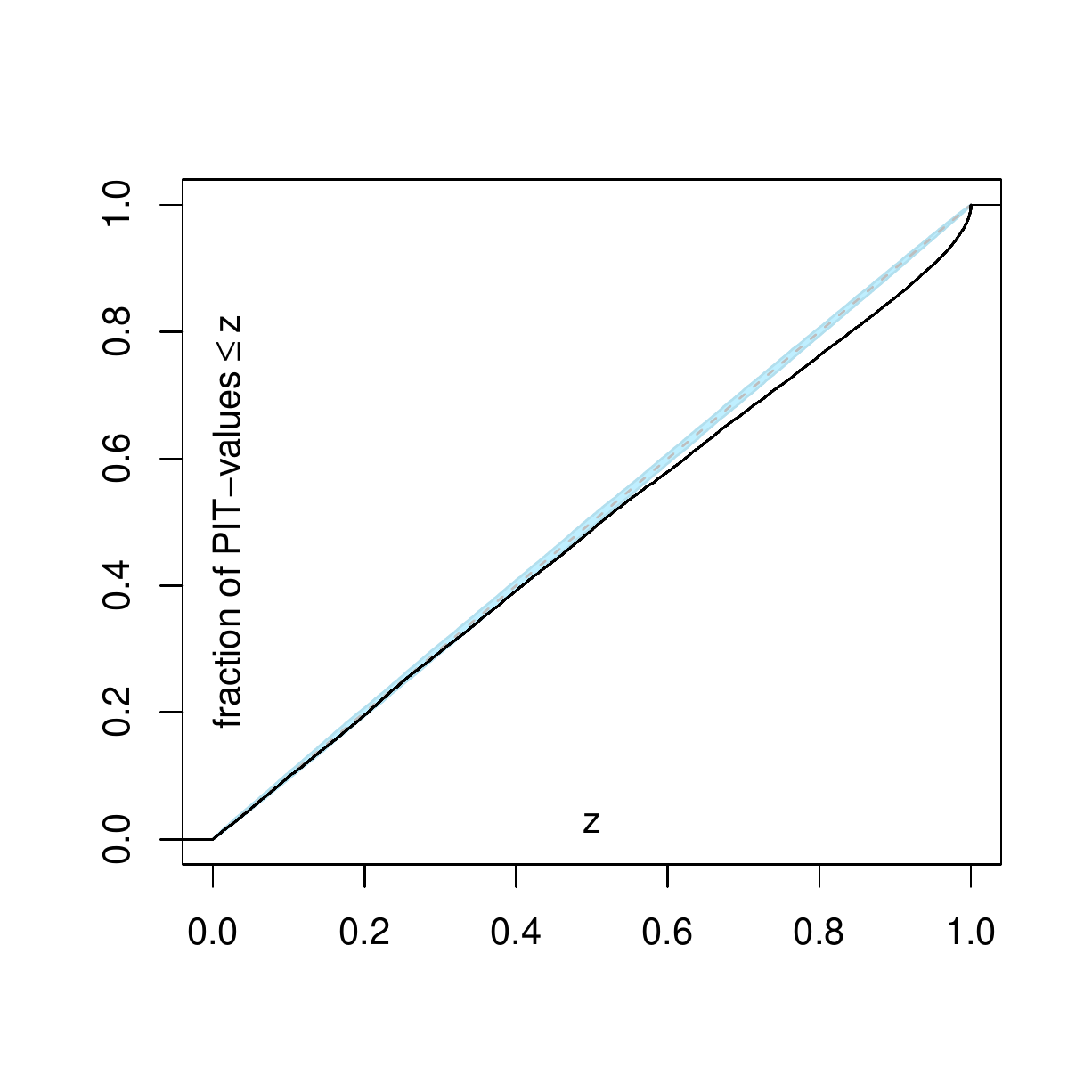}
    \raisebox{0.4cm}{\includegraphics[width=0.31\textwidth]{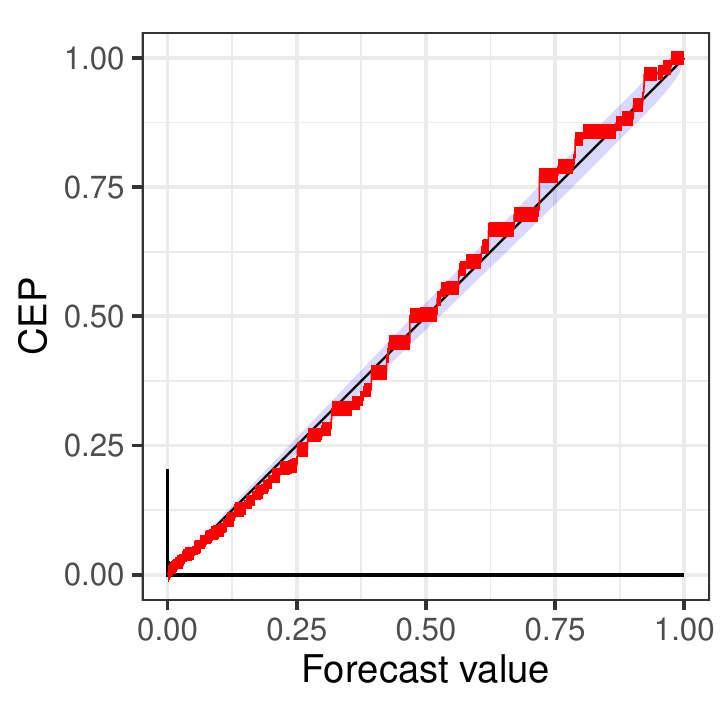}}
    \raisebox{0.9cm}{\includegraphics[width=0.29\textwidth]{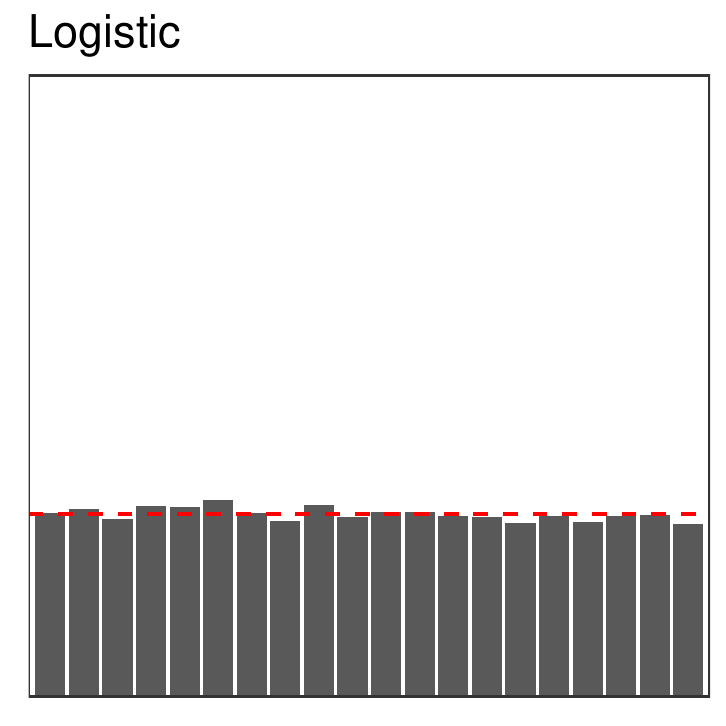}}
    \includegraphics[width=0.38\textwidth]{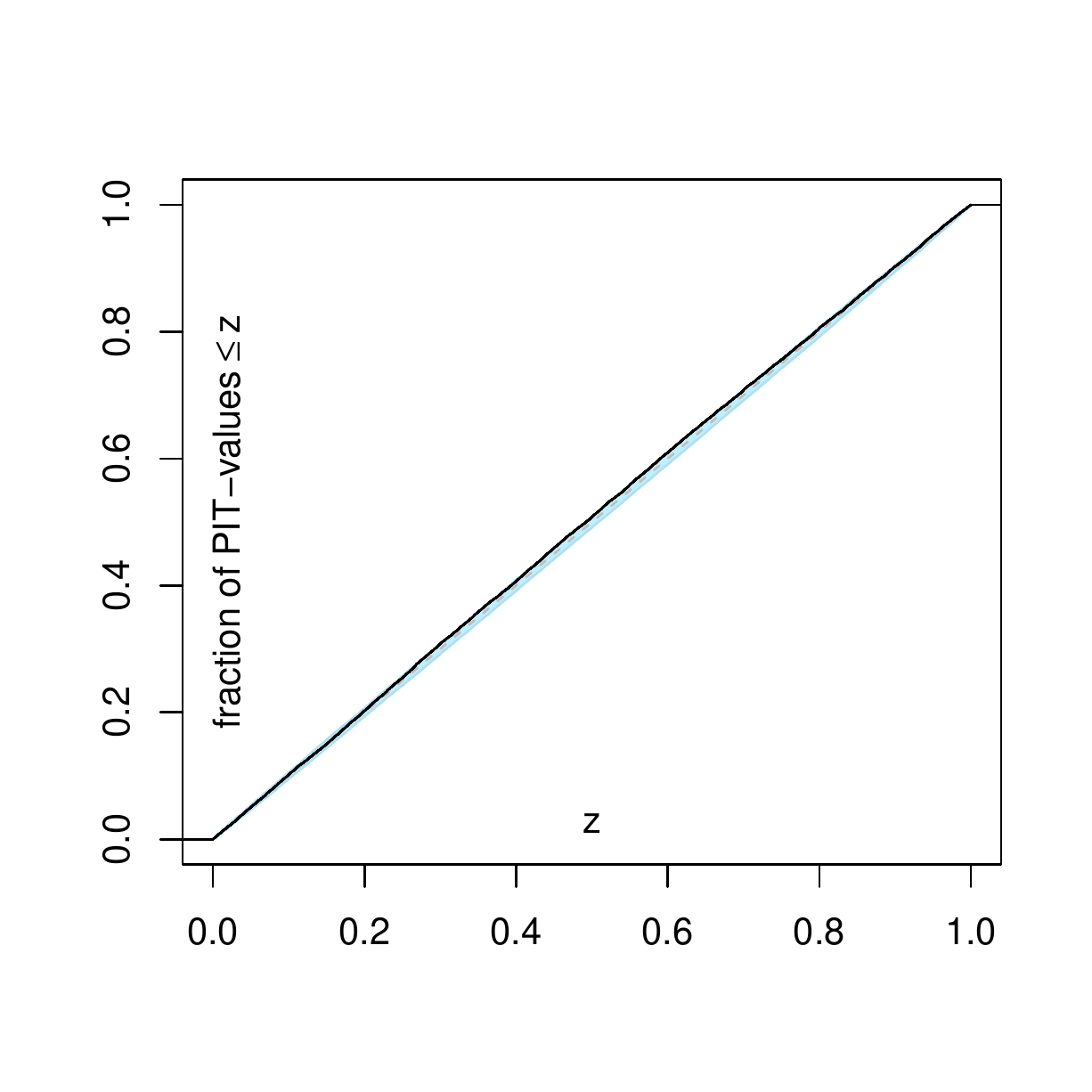}
    \raisebox{0.4cm}{\includegraphics[width=0.31\textwidth]{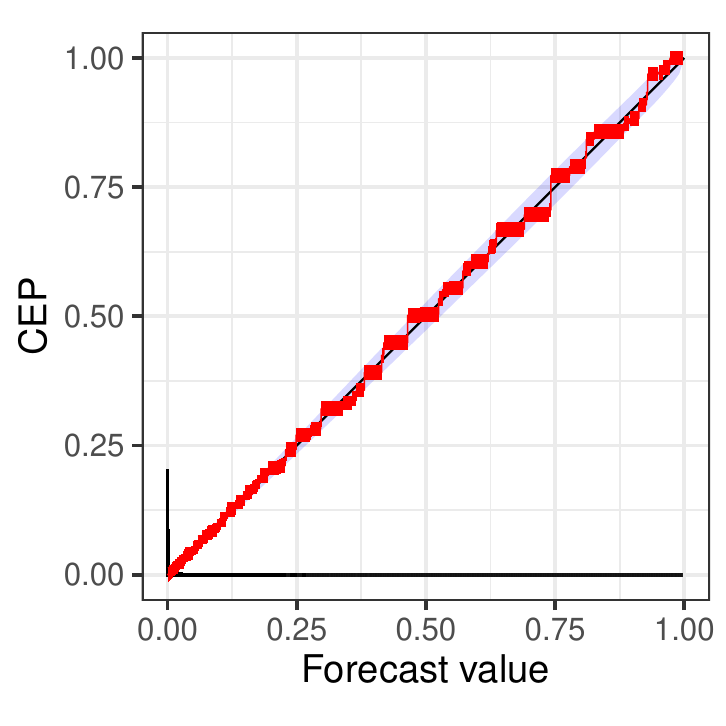}}
    \raisebox{0.9cm}{\includegraphics[width=0.29\textwidth]{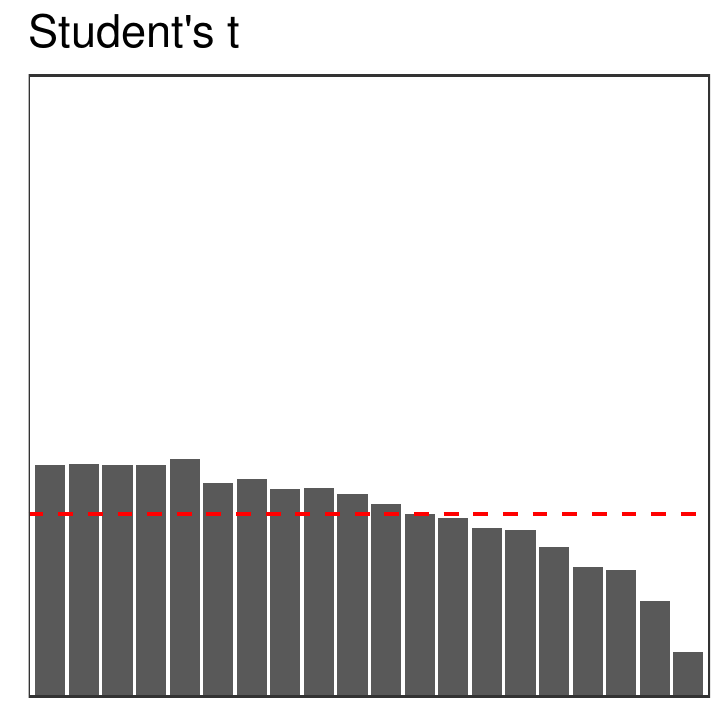}}
    \includegraphics[width=0.38\textwidth]{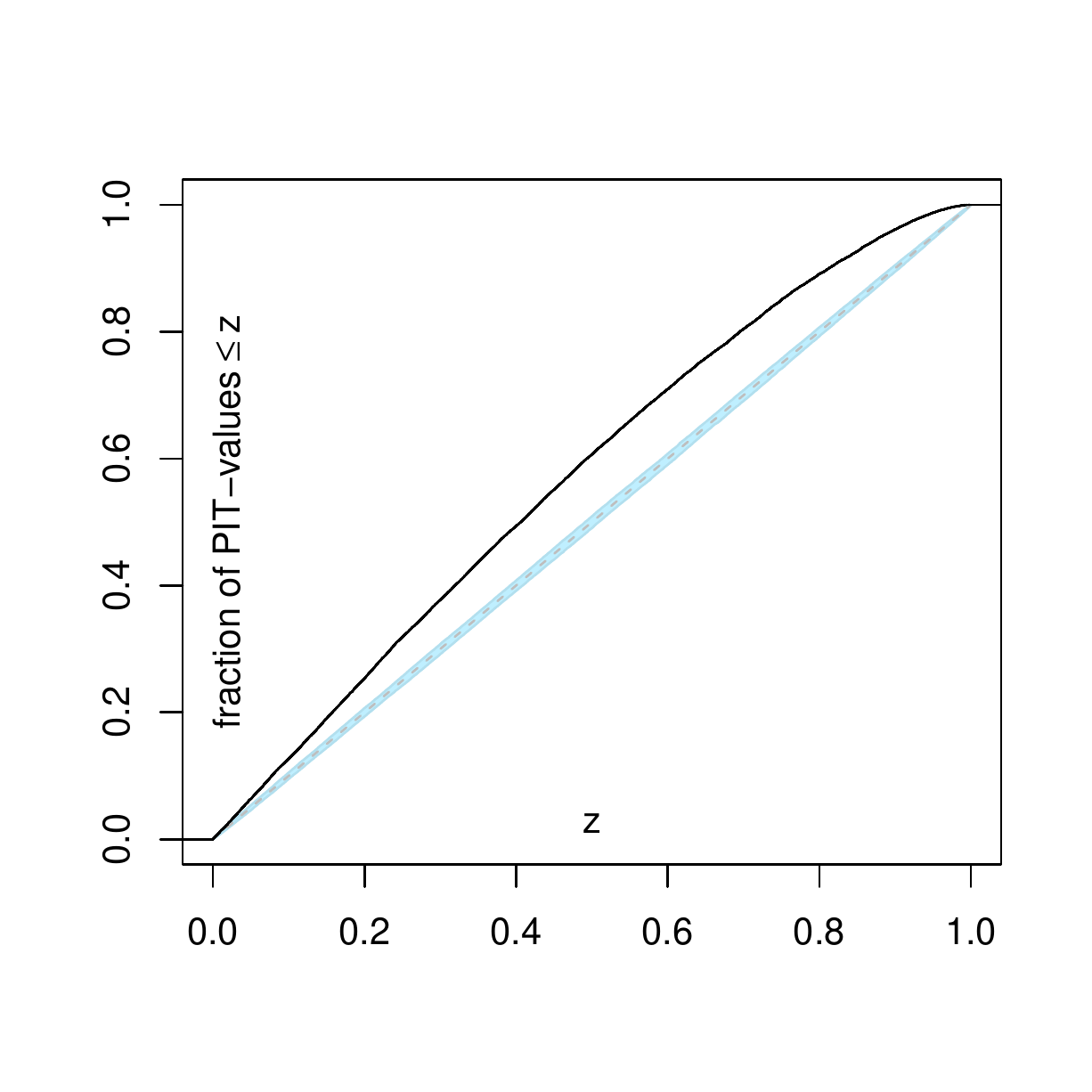}
    \raisebox{0.4cm}{\includegraphics[width=0.31\textwidth]{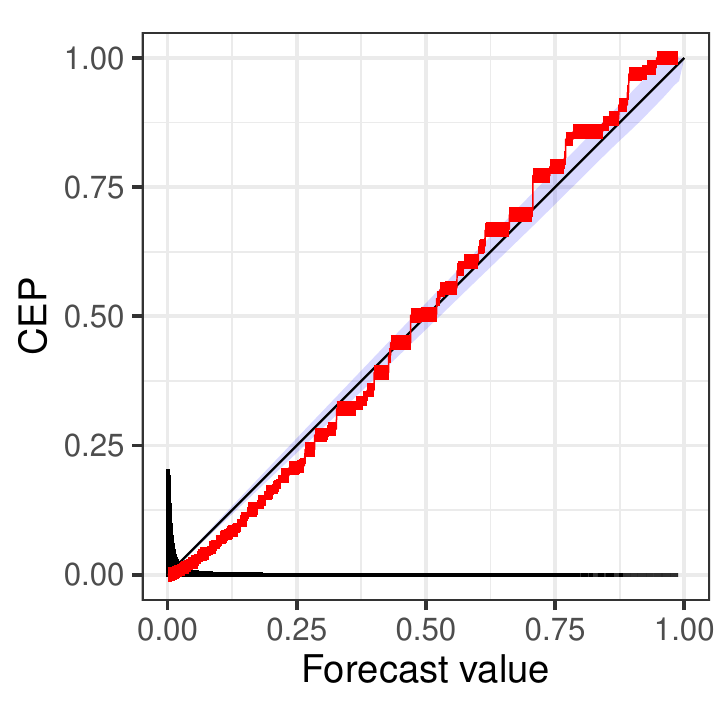}}
    \vspace*{-1cm}
    \caption{Conditional PIT histograms (left) and conditional PIT reliability diagrams (middle) for forecast distributions with light (Normal), perfect (Logistic), and heavy (Student's $t$) tails. The histograms have been constructed using 1,000,000 observations from a logistic distribution, roughly 25,000 of which exceed the threshold $ t = 2 $. Standard reliability diagrams (right) also show the conditional event probabilities (CEP) given the forecast probability that the threshold will be exceeded. The blue shaded regions on the reliability diagrams are consistency intervals, constructed such that a calibrated prediction system would lie within these intervals 99\% of the time.}
    \label{fig:PIT_examples}
\end{figure}

If the cPIT histogram is not uniform, then the shape of the histogram can be used to infer what errors are present in the forecast. For example, suppose the observations are drawn from a logistic distribution with a random mean and fixed variance, and consider three competing forecasters: the first forecaster issues the normal distribution as a forecast, the second issues the logistic distribution, and the final forecaster issues the Student's $ t $ distribution with five degrees of freedom, all of which are constructed to have the same mean and variance as the outcome distribution. Figure \ref{fig:PIT_examples} displays the cPIT histograms for the three approaches, with a threshold equal to two. The logistic forecaster is the ideal forecaster, resulting in a uniform cPIT histogram, whereas the other two forecasters are oppositely biased: the normally distributed forecasts exhibit too light a tail, indicating a large proportion of the observations that exceed $ t $ fall in the tail of the conditional Gaussian distribution, while the Student's $ t $ distribution has a heavier tail than the logistic distribution, resulting in forecasts that over-predict the severity of extreme events. 

\bigskip

While the number of bins to display in a PIT histogram is often chosen to equal the number of possible ranks within a reference ensemble prediction system, there is no canonical choice for the number of bins in a cPIT histogram. Hence, although histogram-based diagnostic tools are commonly employed to assess forecast calibration, we instead recommend visualising conditional calibration using PIT reliability diagrams \citep{Gneiting2022}. PIT reliability diagrams display the empirical cumulative distribution function of the observed PIT values, and, as with standard reliability diagrams, a straight line along the graph's diagonal is indicative of a calibrated prediction system. Conditional PIT reliability diagrams analogously display the conditional PIT values, and cPIT reliability diagrams for the three forecasters in the previous example are presented in Figure \ref{fig:PIT_examples}.

\bigskip

Regardless of how the conditional calibration is visualised, by considering only the outcomes that exceed a threshold, these conditional diagnostic tools inherit some of the disadvantages associated with the outcome-weighted scores that were discussed previously. In particular, the outcome-weighted CRPS only assesses the shape of the conditional distribution, and does not consider the probability of a high-impact event occurring. This is also true for cPIT histograms and cPIT reliability diagrams, meaning they only evaluate the predicted severity of the high-impact event, and not the probability of occurrence. We therefore recommend that they are accompanied by a standard reliability diagram that separately assesses how well the forecasts can predict the occurrence of a high-impact event - akin to how \cite{Holzmann2017} suggest complementing the owCRPS with a scoring rule for binary events. An illustration of this is presented in Figure \ref{fig:PIT_examples} for the Gaussian, logistic, and Student's $ t $ forecasters. These reliability diagrams, constructed using the CORP approach proposed recently by \cite{Dimitriadis2021}, highlight that, despite the differences when predicting event severity, the calibration of the three forecasters does not vary much when predicting the occurrence of a threshold exceedance.

\bigskip

Another disadvantage of the outcome-weighted CRPS is that it cannot easily be applied to ensemble forecasts when interest is on rare events, since the conditional forecast distribution is not always well-defined. Again, this also applies to checks for conditional calibration. As with the owCRPS, this could be addressed by smoothing the ensemble before assessing the calibration. However, generally speaking, we only advise employing checks for conditional calibration to ensemble forecasts when at least a reasonable number of ensemble members (say, 10) are expected to exceed the threshold of interest. While this limits their utility when evaluating ensemble forecasts whilst targeting high-impact events, cPIT histograms and cPIT reliability diagrams could still be useful when assessing the conditional calibration of ensemble forecasts relative to more moderate thresholds: for example, when interest is on precipitation accumulations that exceed zero.

\bigskip

Nonetheless, cPIT histograms and reliability diagrams provide a convenient and easily interpretable graphical approach to visualise calibration conditional on a high-impact event having occurred. While the interpretation of these diagnostic checks is similar to that for conventional checks for overall forecast calibration, formally testing whether a forecast is conditionally calibrated is less straightforward. In particular, the number of observations that exceed the threshold of interest is random and depends on the observed outcomes, rendering standard one-sample tests of uniformity invalid. Instead, more involved statistical tests are required that test for equality of conditional distributions, such as those commonly applied in the field of extreme value theory \citep{Coles2001}.

\bigskip

Throughout this section, the discussion has focused on high-impact events defined as the exceedance of a relevant threshold, i.e.\ corresponding to a weight function $ w(z) = \one\{ z > t \}$. In theory, cPIT histograms and reliability diagrams could be extended to more general weight functions. This would require identifying the random variable that follows the weighted distribution $ F_{w} $ in Equation \ref{eq:weightedF}. This will change for each weight function being considered, and is, in general, not a trivial task. However, we note that the weighted distribution $ F_{w} $ is generally not easy to interpret, and hence, even if we were able to construct the weighted PIT histogram corresponding to a general weight function, it would not be straightforward to use the resulting histogram to diagnose exactly what errors are present in the prediction system. For this reason, we restrict further attention to the cPIT histograms and reliability diagrams introduced above.
 
\bigskip

\subsubsection*{Multivariate conditional PIT histograms}

Just as weighted scoring rules can be designed to target multivariate outcomes during forecast evaluation, the cPIT histograms introduced in the previous section can also be extended to the multivariate case. There is no canonical definition of a multivariate rank or PIT histogram, and several contrasting approaches have been proposed to construct them \citep[see][]{Thorarinsdottir2018book}. The general approach, as outlined by \cite{Ziegel2017}, is to define a pre-rank function, which condenses the multivariate forecasts and observations to univariate objects, and then to assess the calibration of these transformed forecasts using standard univariate rank or PIT histograms. The various approaches that have been proposed differ in their choice of pre-rank function.

\bigskip

Regardless of the chosen pre-rank function, we can straightforwardly adapt this approach to construct multivariate cPIT histograms that emphasise particular multivariate outcomes when assessing forecast calibration. For example, consider a multivariate threshold of interest, $ t \in \mathbbm{R}^{d} $, and suppose we are interested in instances where the threshold is exceeded along all dimensions. By applying the pre-rank function to this multivariate threshold, we can obtain a univariate threshold. Note that this univariate threshold will change for each forecast case, since the pre-rank function typically depends on the forecast. Nonetheless, having obtained a univariate threshold, a cPIT histogram or cPIT reliability diagram can now easily be constructed as described in the previous section.

\bigskip

Note, however, that the challenges mentioned previously when assessing the conditional calibration of ensemble forecasts also apply here, and, since multivariate weather forecasts are more regularly in the form of finite ensembles, these issues will be yet more prevalent in the multivariate setting. In the case study presented in the following section, the multivariate forecasts are all ensemble forecasts, and hence we do not employ this approach to assess the conditional calibration of the multivariate forecasts. 

\bigskip

\section{Case study: evaluating heatwave forecasts}
\label{section:casestudy}

\subsection{Extreme heat events}


The verification techniques discussed in the previous section provide a means of evaluating forecasts with respect to high-impact events. In this section, we demonstrate the practical benefit afforded by these techniques by using them to evaluate operational weather forecasts for heatwaves and extreme heat events. The impacts associated with high temperatures can vary depending on several meteorological and societal factors, and there is thus no universal definition of a heatwave or dangerous heat event. Since these impacts can be mitigated through effective warning systems, we define heat events using operational heat warning criteria adopted by the Swiss Federal Office of Meteorology and Climatology (MeteoSwiss), determined following a recent study on how high temperatures affect human health in Switzerland \citep{Ragettli2017}.

\bigskip


MeteoSwiss issue heat warnings of three different levels, with a higher level associated with a higher impact. All heat levels are defined in terms of the daily mean temperature over a three day period, as summarised in Table \ref{tab:WarningLevels}. For completeness, Table \ref{tab:WarningLevels} also includes a level one heat event, synonymous with the occurrence of low or moderate temperatures; the four levels therefore comprise an exhaustive set of the possible daily mean temperatures over three days. As expected, non-dangerous heat occurs on the vast majority (97\%) of instances, whereas the most severe heat level occurs just 0.04\% of the time. The MeteoSwiss warning levels do not change depending on the location, thereby assuming that the dangers associated with heat events do not vary substantially within the relatively small country of Switzerland. 

\bigskip

\begin{table}
    \centering
    \begin{tabular}{| c | c | c |}
    \hline
    Heat level & Criterion & Rel. Freq. (\%) \\
    \hline
    1 & T $< 25^{\circ}$C on all three days & 97.12 \\
    2 & T $\geq 25^{\circ}$C on one or two days & 2.40 \\
    3 & T $\geq 25^{\circ}$C on all three days, T $< 27^{\circ}$C on at least one day & 0.45 \\
    4 & T $\geq 27^{\circ}$C on all three days & 0.04 \\
    \hline
    \end{tabular}
    \caption{MeteoSwiss heat warning levels given daily mean temperatures (T) over a three day period, and the relative frequency with which each level occurs in the data under consideration.}
    \label{tab:WarningLevels}
\end{table}

\subsection{Data}

Since these heat event definitions depend only on the daily mean temperature, we study forecasts for this weather variable. In particular, we consider daily mean temperature forecasts obtained from an operational ensemble prediction system at MeteoSwiss, which is based on a high-resolution numerical weather prediction (NWP) model from the Consortium for Small-Scale Modeling (COSMO-E). The COSMO-E model operates at a horizontal resolution of 2.2km over Switzerland and the surrounding area, and produces ensemble forecasts comprised of 21 members, all of which are initialised at 00 UTC in this study. Further details regarding COSMO-E are provided by \cite{Keller2021} and references therein.

\bigskip

However, even high-resolution NWP models are unable to resolve Switzerland's complex topography, leading to large temperature biases on valley-floors and mountain-tops. To account for this, a simple lapse-rate bias correction is added to the COSMO-E forecasts, which takes into account the difference between the height of the model at each location and the true altitude; we assume a constant lapse-rate of 0.6 degrees Celsius per 100m.

\bigskip

\begin{figure}
    \centering
    \includegraphics[width=0.5\linewidth]{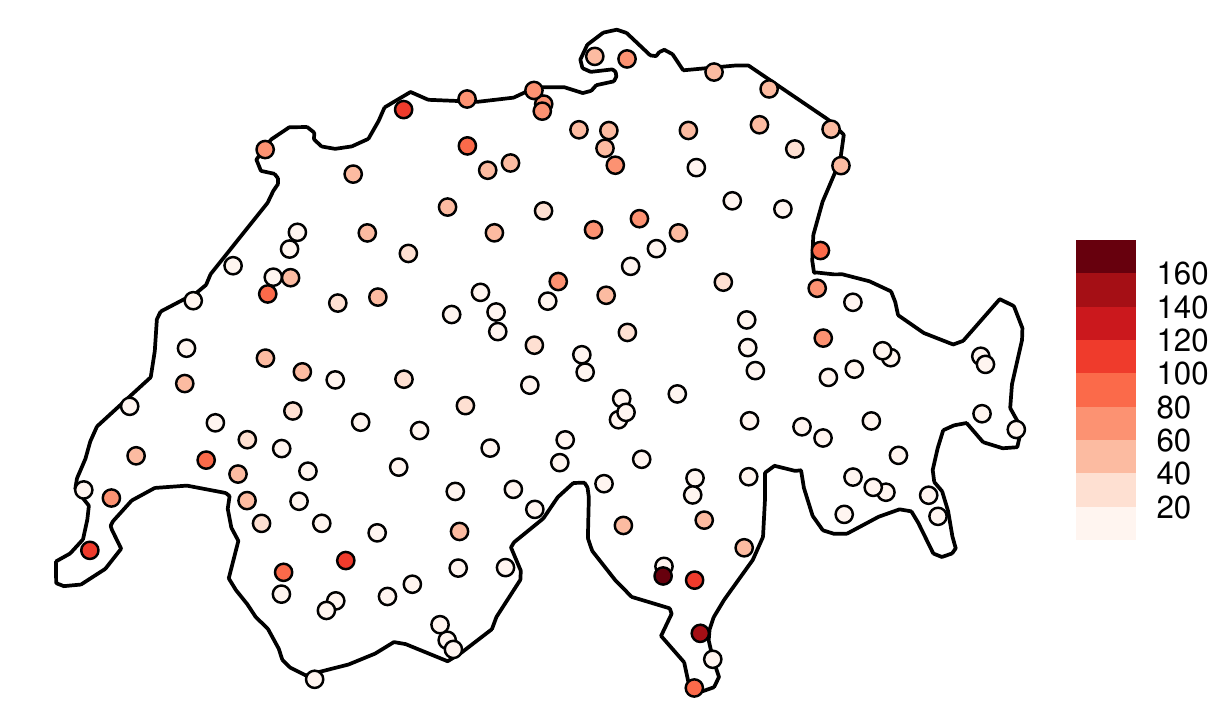}
    \caption{Frequency of heat events of level two or greater at each station of interest, over all seven summer seasons under consideration.}
    \label{fig:WarnFreq}
\end{figure}

The forecasts are assessed against observational temperature records at 149 weather stations across Switzerland, with the gridded COSMO-E output interpolated to individual stations using a nearest grid-point approach. These stations are all operated by MeteoSwiss and subject to rigorous quality control procedures. The stations are displayed in Figure \ref{fig:WarnFreq}, along with the number of extreme heat events (i.e.\ level two or greater) that occur at each station during the period of interest. Although there are several stations at which an extreme heat event does not occur, all stations are utilised in the subsequent analysis since forecasts should also be assessed in their ability to predict when an extreme event will not occur. 

\bigskip

Forecasts and observations are available for the seven year period between 2014 and 2020, and we restrict attention to extended summer seasons (May-September) in order to focus on extreme heat. This results in roughly 150,000 forecast-observation pairs to analyse at each forecast lead time. The COSMO-E forecasts extend out to five days, but since the heatwaves are defined over a three day period, forecasts are only considered over the coming three days. The forecasts are evaluated at each lead time separately using univariate verification techniques, while multivariate tools are used to assess the forecasts over the entire three day period. In doing so, forecasts can be evaluated in their ability to predict the temporal evolution of the daily mean temperature, which is key when focus is on heatwaves. Furthermore, while we evaluate short-range forecasts for these heat events, we note that the techniques outlined in the following sections can readily be applied to longer-range forecasts. 

\bigskip

\subsection{Statistical post-processing}

State-of-the-art ensemble prediction systems typically exhibit systematic biases when forecasting surface weather variables. To remove these biases and re-calibrate the ensemble output, statistical post-processing is typically applied to the forecasts \citep[see][for a review]{Vannitsem2018}. We re-calibrate the COSMO-E daily mean temperature forecasts using the ensemble model output statistics (EMOS) framework proposed by \cite{Gneiting2005}. EMOS assumes that the variable to be forecast follows a certain parametric distribution, whose moments depend linearly on those of the corresponding ensemble forecast. We assume here that the daily mean temperature at a given time and location is normally distributed. 

\bigskip

To account for local structures within the COSMO-E forecast biases, two additional predictors are incorporated into the post-processing model: a topographic position index (TPI) that reflects the change in elevation between a station and those in a local neighbourhood of 2km radius, and a measure of the height difference between the COSMO-E model and reality (MHD). The inclusion of these two spatial covariates follows other recent studies on the post-processing of COSMO-E temperature forecasts in Switzerland \citep[e.g.][]{Keller2021}. These additional predictors allow the model to account for local features in the forecast biases despite fitting a single post-processing model simultaneously to forecasts at all stations. 

\bigskip

The post-processing model can be formalised as follows. Let $ Y $ denote the daily mean temperature at a given station, time, and lead time, and let $ \bar{x} $ and $ v $ denote the mean and variance of the corresponding COSMO-E ensemble members, respectively. Then, the model assumes that
\begin{equation}\label{eq:pp_model}
    Y = \beta_{0} + \beta_{1}\bar{x} + \beta_{2} \mathrm{MHD} + \beta_{3} \mathrm{TPI} + \epsilon, \hspace{1cm} \epsilon \sim \mathcal{N}(0, \sigma_{0} + \sigma_{1} v),
\end{equation}
where $ \mathcal{N}( \mu, \gamma) $ denotes the normal distribution with mean $ \mu $ and variance $ \gamma $. Note that the TPI and MHD depend on the station under consideration, but not on the time or lead time. The variance of this model could similarly be set up to depend on the MHD and TPI, but this was not found to provide much benefit. 

\bigskip

The post-processing model parameters $ \beta_{0}, \beta_{1}, \beta_{2}, \beta_{3}, \sigma_{0}, \sigma_{1} $ link the predictors to the observations. A separate set of parameters is estimated for each forecast lead time, thereby acknowledging that the relationship between the forecast and the observation will change as the forecast horizon increases. As in \cite{Keller2021}, the parameters are estimated by minimising the CRPS over a rolling training window containing the previous 45 forecast-observation pairs, allowing the model to also account for recent patterns in the forecast biases. 

\bigskip

Post-processing is applied to the daily mean temperature forecast at each lead time separately. Since dangerous heat events are often a multivariate phenomenon, we use copulas to convert these individual forecast distributions into a temporally coherent multivariate forecast over the three day period. To do so, we employ ensemble copula coupling \citep[ECC;][]{Schefzik2013}, an empirical copula-based approach. ECC works by converting the univariate post-processed forecast distributions at each lead time to an ensemble forecast, by selecting 21 evenly-spaced quantiles from each distribution, before reordering the resulting ensemble members so that the rankings of the ensemble members at each lead time are the same as in the corresponding COSMO-E ensemble. 

\bigskip

By comparing the performance of this post-processing model to the raw COSMO-E output, we can investigate how post-processing affects predictions of high-impact events: \cite{Pantillon2018}, among others, have recently postulated that post-processing can hinder forecasts of extreme events due to a regression-to-the-mean type effect. We additionally compare the  COSMO-E and post-processed forecasts to a climatological prediction. The climatological forecast again assumes that the temperature is normally distributed, but no predictors are employed within this distribution. The mean and variance of this climatological distribution are estimated over a 45-day rolling window, similarly to the post-processing model, though a separate climatology is estimated for each station separately to incorporate local information. An empirical copula is then applied to the climatological forecasts to generate a coherent multivariate forecast. 

\bigskip

\subsection{Results}

\subsubsection*{Overall forecast performance}

The accuracy of the three prediction systems is assessed here using the CRPS. The scores for the three methods, averaged over all forecast cases and stations, are displayed in Table \ref{tab:Scores}. As expected, the climatological forecast performs considerably worse than the approaches that utilise the COSMO-E output, while post-processing offers improvements of around 15\% upon the raw model output at all lead times.

\bigskip

The energy score and variogram score are also presented in Table \ref{tab:Scores}. The climatological forecasts again perform considerably worse than the two other methods, while the post-processed forecasts significantly outperform the COSMO-E ensembles when assessed using the energy score, with a relative improvement similar to that obtained from the CRPS. However, post-processing does not provide any benefit with respect to the variogram score. Since the variogram score is more sensitive to the forecast dependence structure, this suggests that the benefit of post-processing is largely due to improvements in the univariate forecast distributions, rather than in the multivariate dependence structure.

\bigskip

\begin{table}[!b]
    \centering
    \begin{tabular}{| c | c c c | c | c |} 
        \hline
          & \multicolumn{3}{| c |}{CRPS} & ES & VS \\
          \cline{2-4}
          & 1 day & 2 days & 3 days & &  \\
          \hline
          Climatology & 2.36 & 2.36 & 2.37 & 4.65 & 4.42 \\
          COSMO & 1.05 & 1.08 & 1.07 & 2.02 & 1.36 \\
          Post-processed & 0.88 & 0.92 & 0.92 & 1.76 & 1.37  \\
         \hline
    \end{tabular}
    \caption{The CRPS at each lead time, as well as the energy score and variogram score for the climatological, COSMO-E, and post-processed forecasts. Scores have been aggregated over all years and stations.
    }
    \label{tab:Scores}
\end{table}

\begin{figure}[t!]
    \centering
    \includegraphics[width=0.48\linewidth]{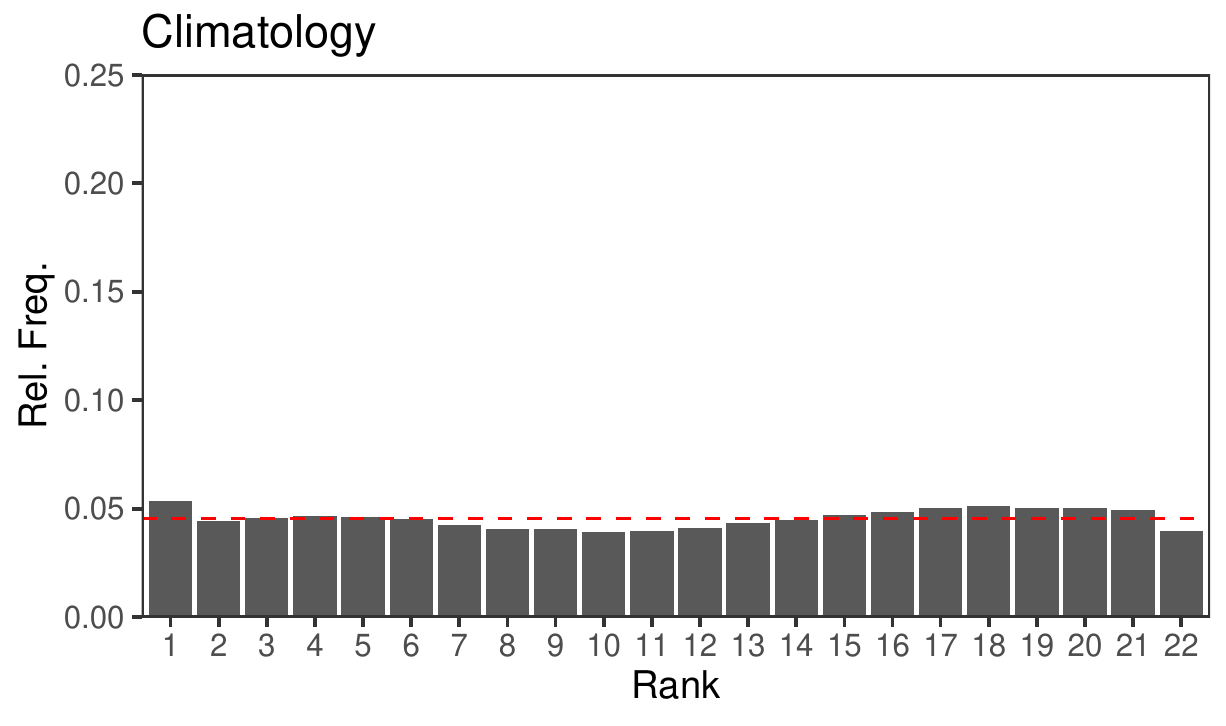}
    \includegraphics[width=0.48\linewidth]{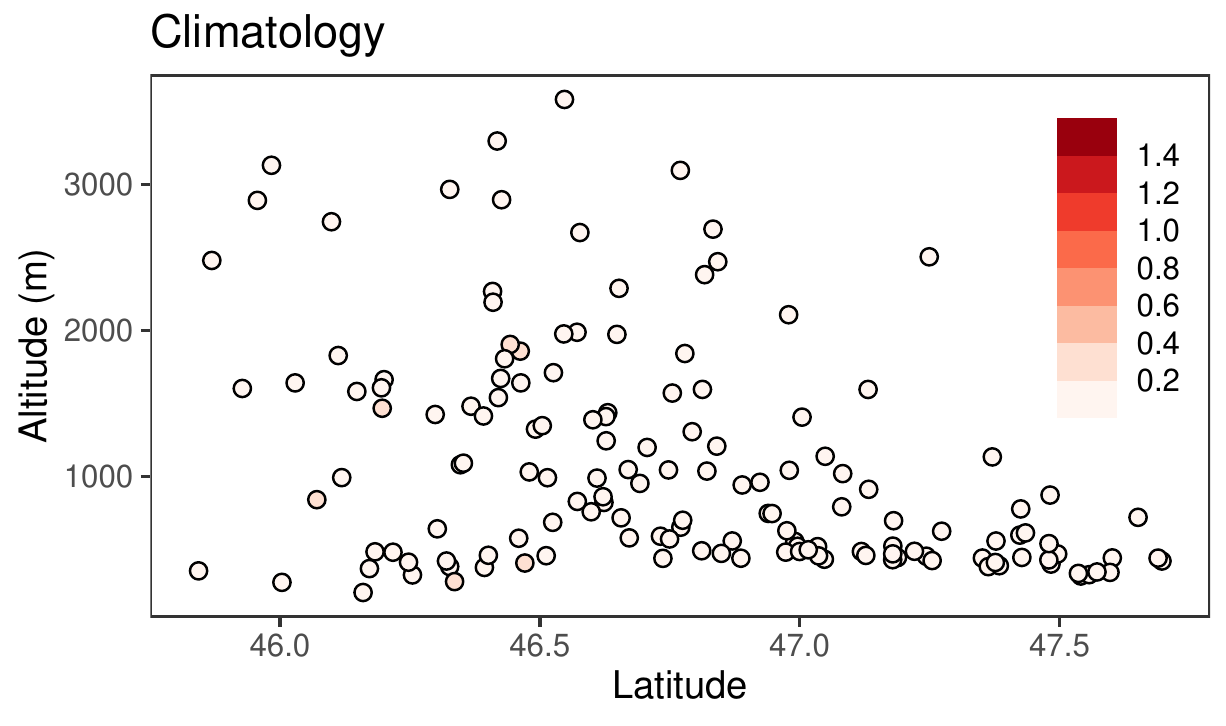}
    \includegraphics[width=0.48\linewidth]{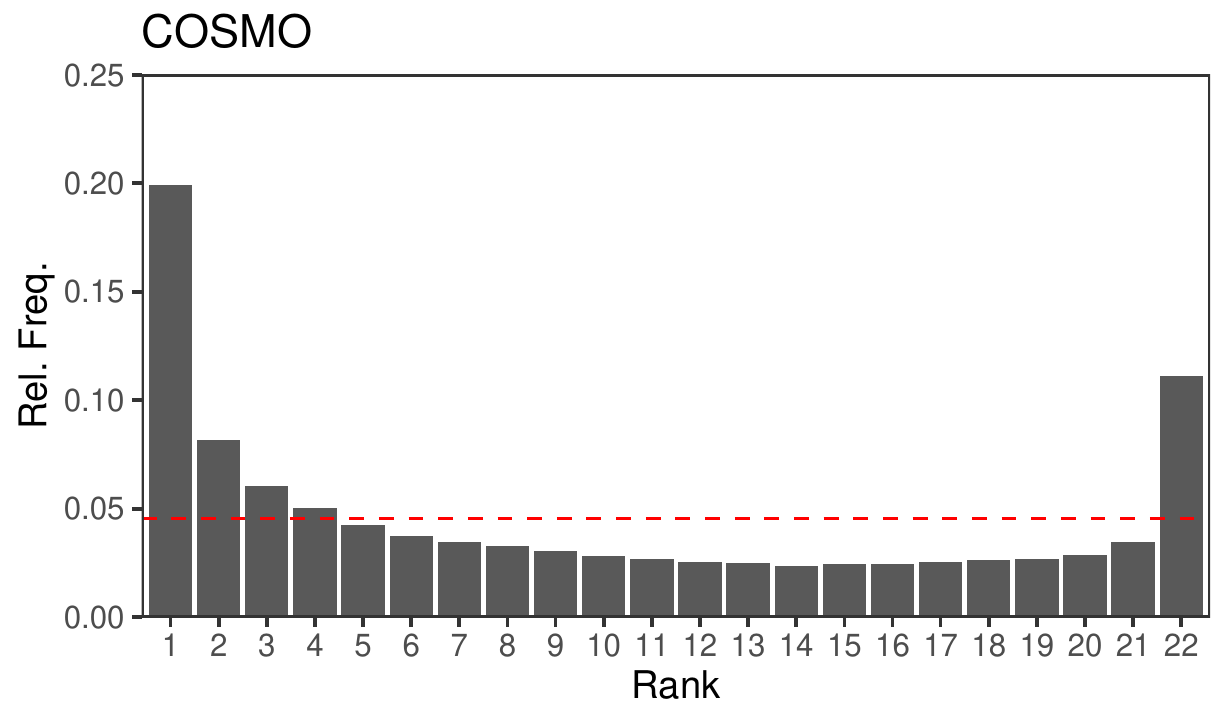}
    \includegraphics[width=0.48\linewidth]{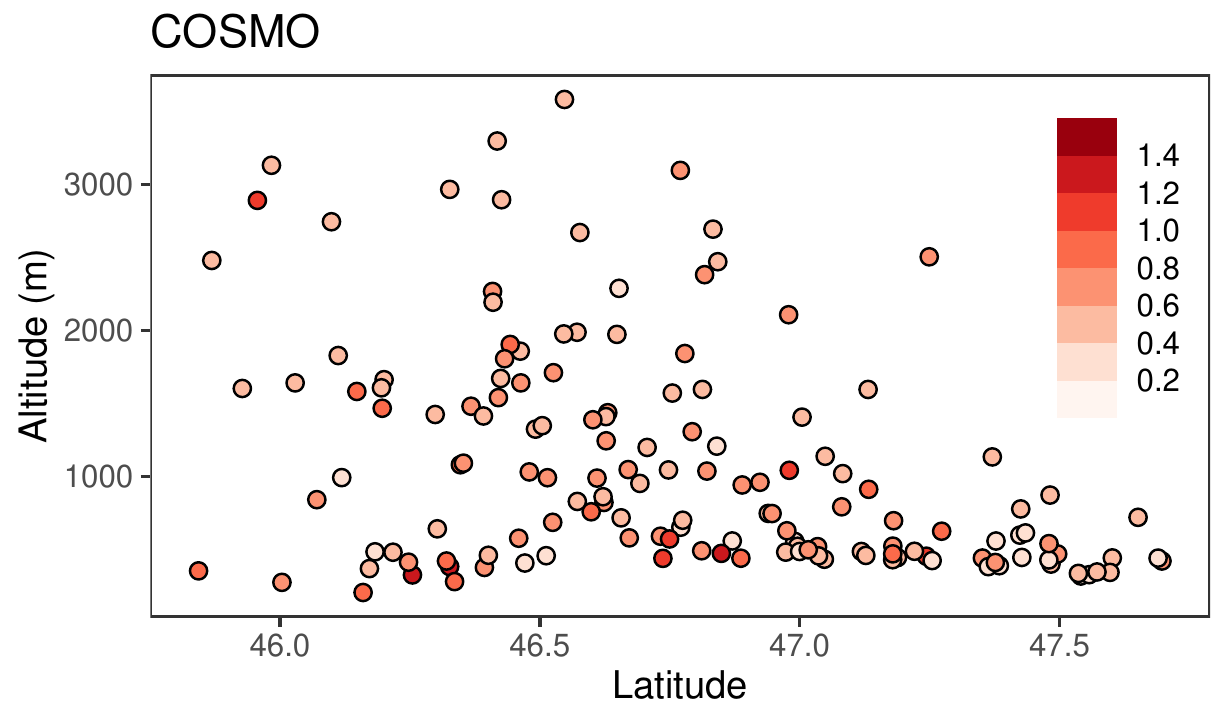}
    \includegraphics[width=0.48\linewidth]{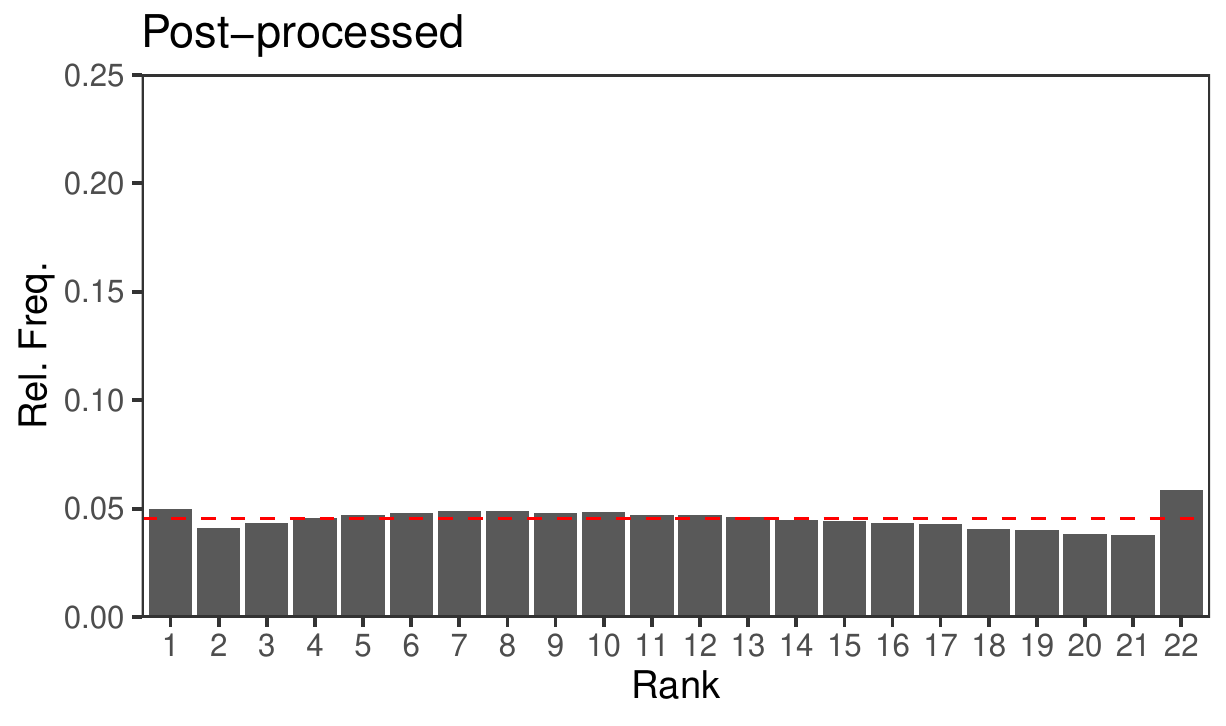}
    \includegraphics[width=0.48\linewidth]{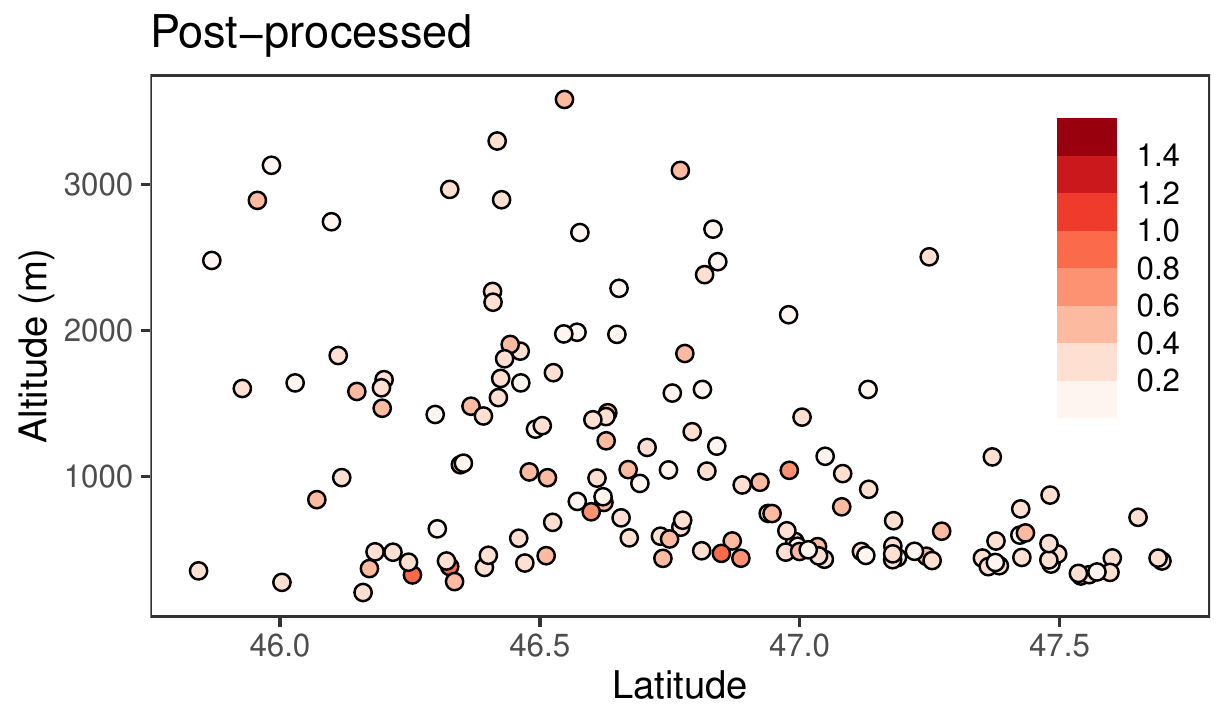}
    \caption{Rank histogram for the COSMO-E ensemble and PIT histograms for the climatological and post-processed forecast distributions at a lead time of three days. The ranks have been aggregated over all years and stations, and the horizontal red line is indicative of perfect calibration. A measure of the miscalibration in the histogram is also shown as a function of the station latitude and altitude for all three methods.} 
    \label{fig:RankHists}
\end{figure}

The calibration of the competing prediction systems is assessed using rank and PIT histograms, which are displayed in Figure \ref{fig:RankHists}. The results are shown at a lead time equal to three days, though similar conclusions are drawn at other lead times. The COSMO-E forecasts are considerably under-dispersed on average, which is commonly the case for operational weather forecasts for surface weather variables, while the post-processing model generates forecasts that are considerably better-calibrated. The climatological forecasts are also well-calibrated, despite performing poorly with respect to both univariate and multivariate scoring rules.

\bigskip

These rank and PIT histograms have been constructed from the forecasts and observations at all stations of interest. However, due to the complex topography of Switzerland, forecast calibration will likely change depending on the location. Figure \ref{fig:RankHists} additionally contains a reliability index corresponding to each of the 149 stations, as a function of the station latitude and altitude (defined as the height above sea level). The reliability index, introduced in \citet[][Equation 1]{DelleMonache2006}, measures the absolute deviation of the bars in the histogram from the uniform red line: the index is therefore minimised at zero, with larger values indicating more severe miscalibration. The reliability index for the COSMO-E forecasts tends to be marginally smaller at higher altitudes than lower altitudes, though the improvement in calibration gained by post-processing appears to be fairly insensitive to the station's location. The climatological forecasts produce yet smaller reliability indices.

\bigskip

\subsubsection*{Predicting heatwave severity}

\begin{figure}[!t]
    \centering
    \includegraphics[width=0.5\linewidth]{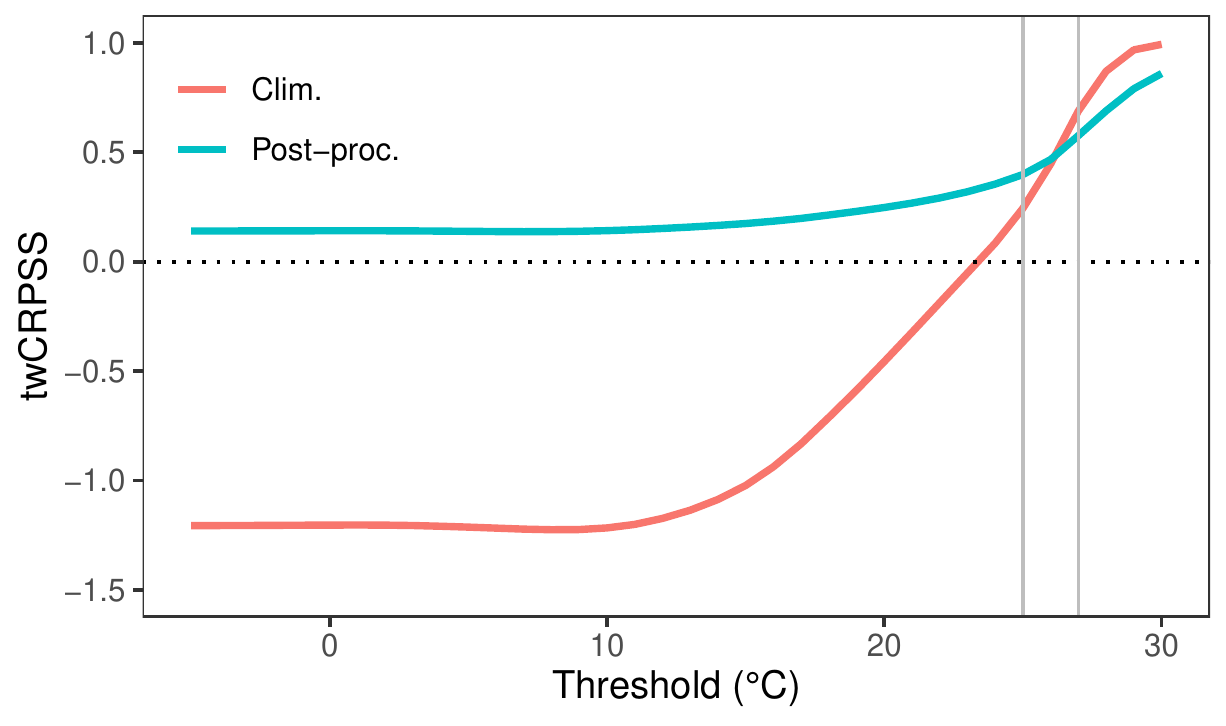}
    \includegraphics[width=0.5\linewidth]{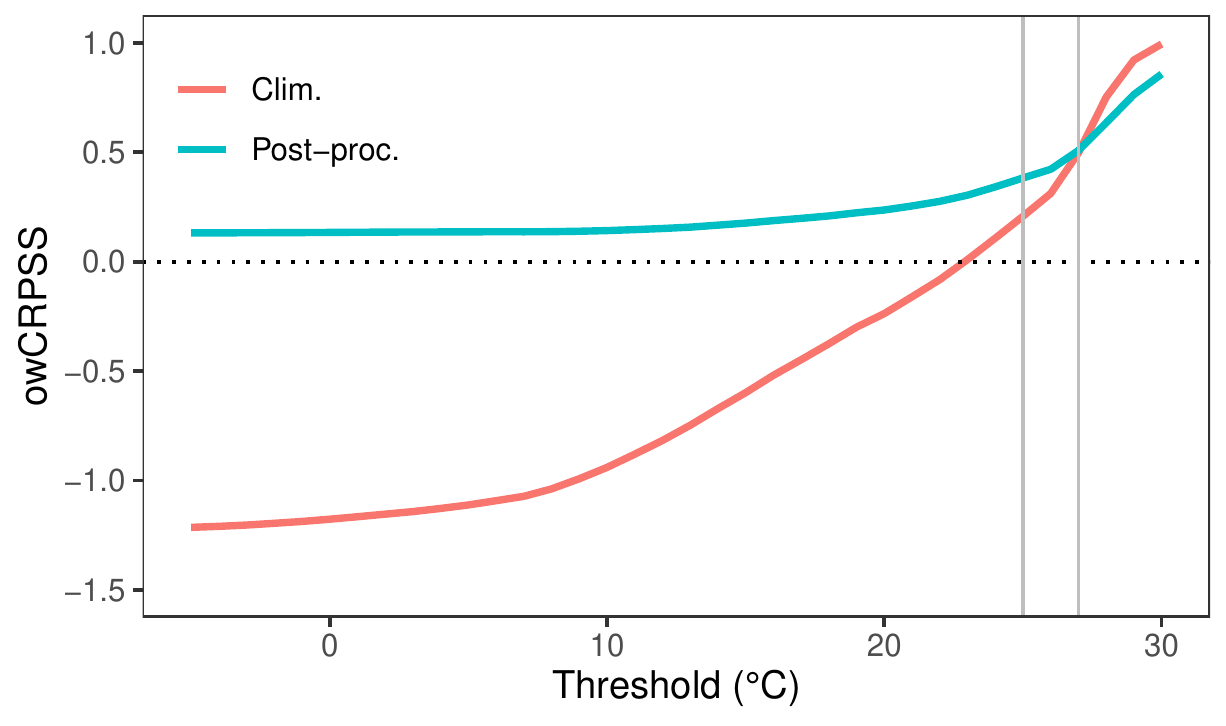}
    \includegraphics[width=0.5\linewidth]{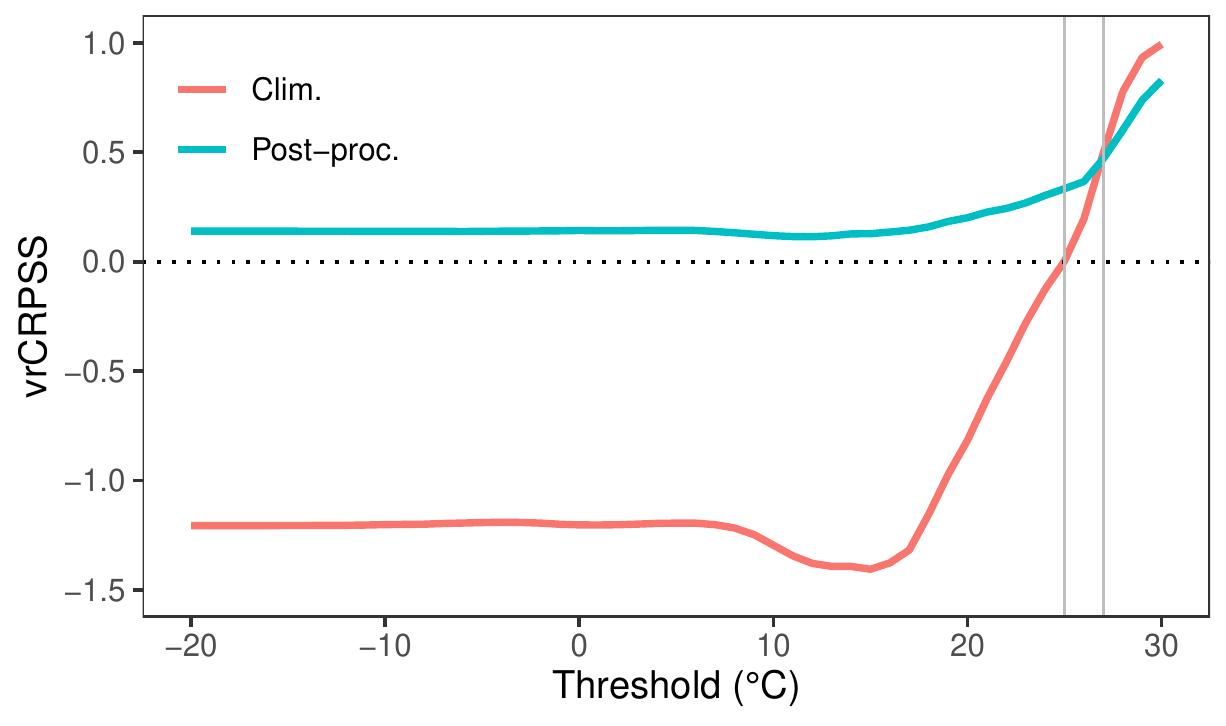}
    \caption{Skill scores for the twCRPS, owCRPS, and vrCRPS as a function of the threshold used in the weight function $ w(z) = \one\{z > t\} $ at a lead time of three days. The skill scores are shown for the climatological and post-processed forecast distributions, with the COSMO-E forecasts as the reference approach. Vertical grey lines are shown at the thresholds $ t = 25 $ and $ t = 27 $.}
    \label{fig:WeightedCRPS}
\end{figure}

In the univariate case, to evaluate how well the forecasts capture the severity of extreme events, the three weighted versions of the CRPS can be employed at each lead time. Figure \ref{fig:WeightedCRPS} displays these scores at a lead time of three days as a function of the threshold employed in the weight function $ w(z) = \one\{z > t\} $, which emphasises events that exceed the threshold $ t $. The owCRPS has been complemented with the Brier score (Equation \ref{eq:owcrps+bs}), and, to ensure this score is well-defined, the COSMO-E ensembles are smoothed using a normal distribution prior to calculating this weighted score. The additional parameter in the vrCRPS is set to $ x_{0} = 0 $. 

\bigskip

The scores are displayed in the form of skill scores, with the raw COSMO-E ensemble forecasts as the reference. A positive skill score indicates an improvement upon the COSMO-E forecasts, whereas a negative skill score suggests the reference forecasts are more accurate.  Vertical lines are displayed at thresholds of 25 and 27 degrees, which correspond to the thresholds employed within the heat event definitions considered here. As the threshold becomes more negative, the weight function tends to one, and all scores should therefore tend to the skill scores obtained from the unweighted CRPS. As expected in this case, the climatological forecasts are significantly worse than the COSMO-E output, while the post-processed forecasts offer improvements of roughly 15\%. 

\bigskip

\begin{figure}[!t]
    \centering
    \includegraphics[width=0.3\linewidth]{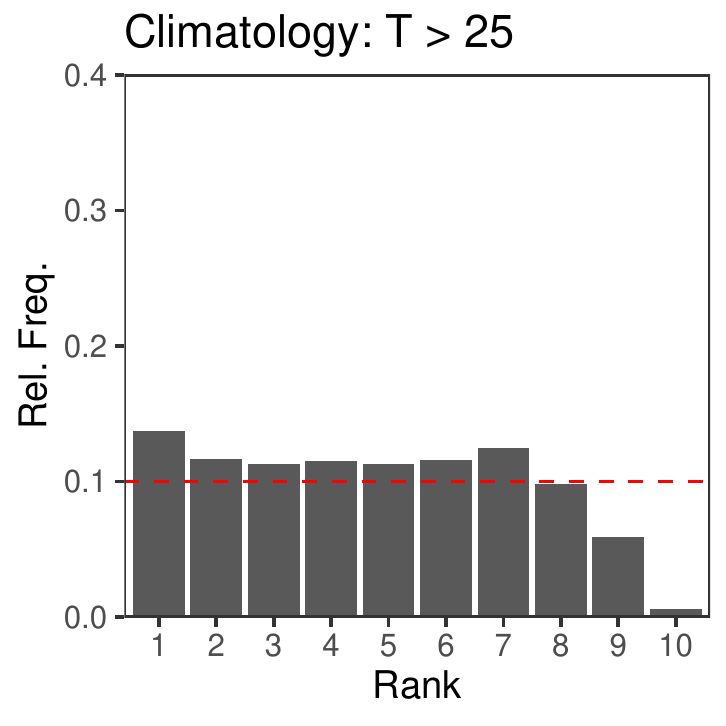}
    \raisebox{-0.5cm}{\includegraphics[width=0.36\linewidth]{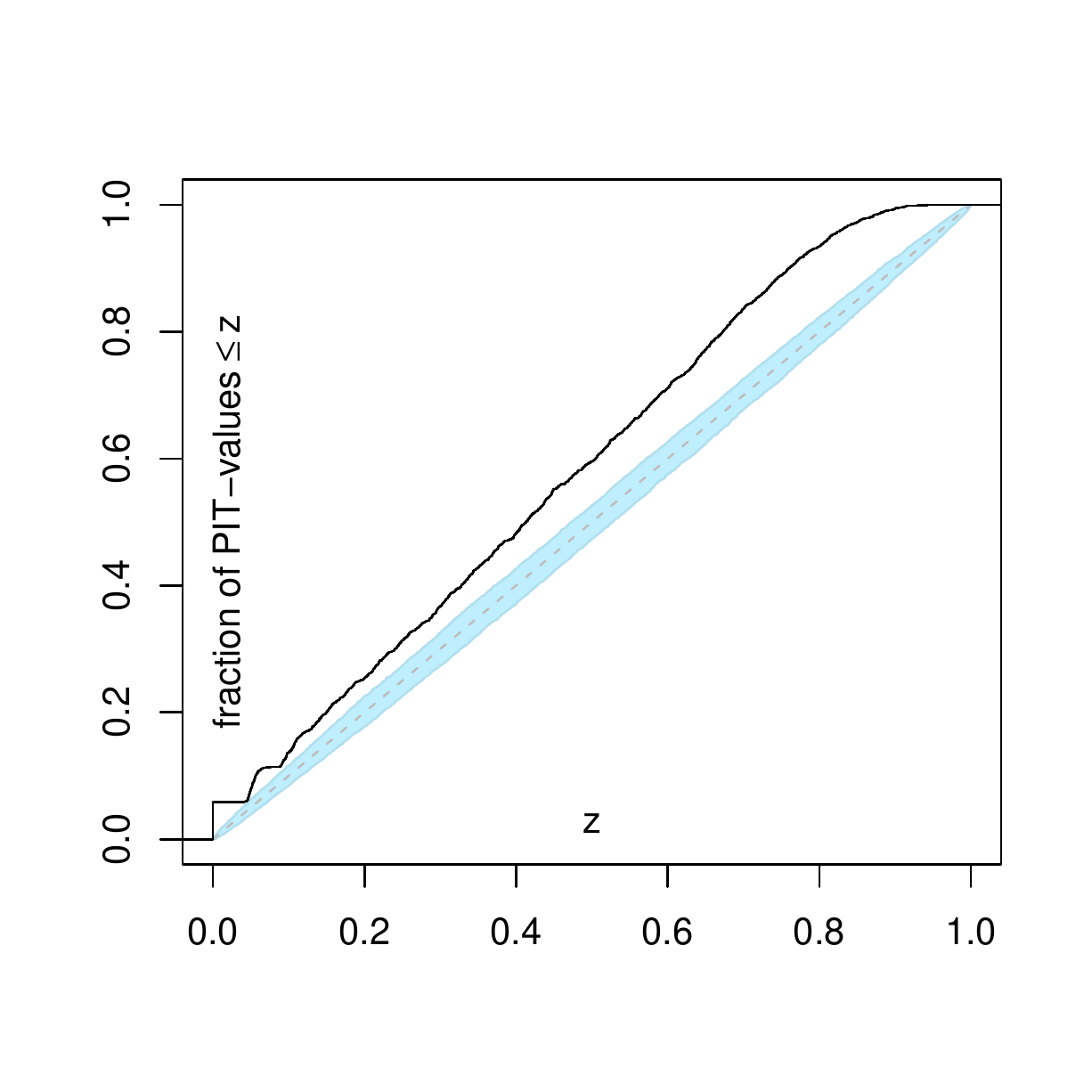}}
    \raisebox{-0.2cm}{\includegraphics[width=0.3\linewidth]{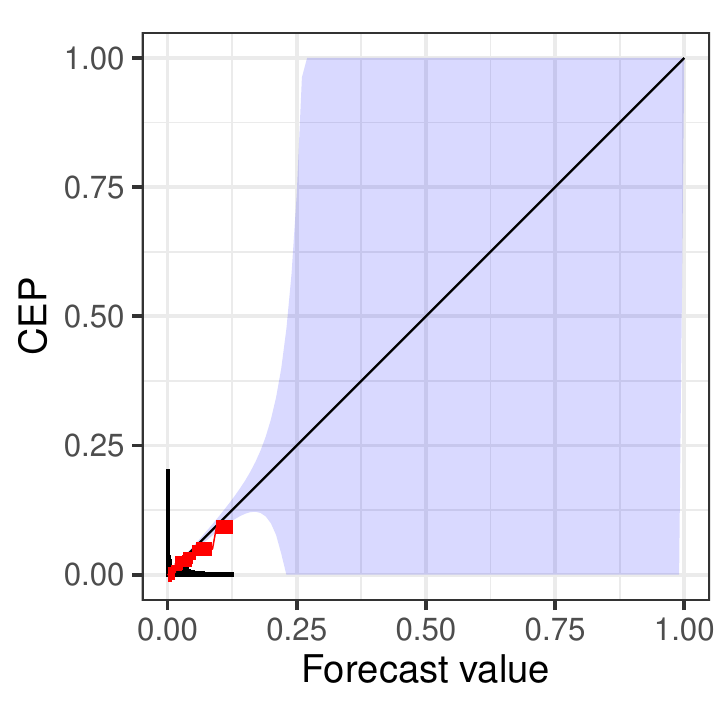}}
    \includegraphics[width=0.3\linewidth]{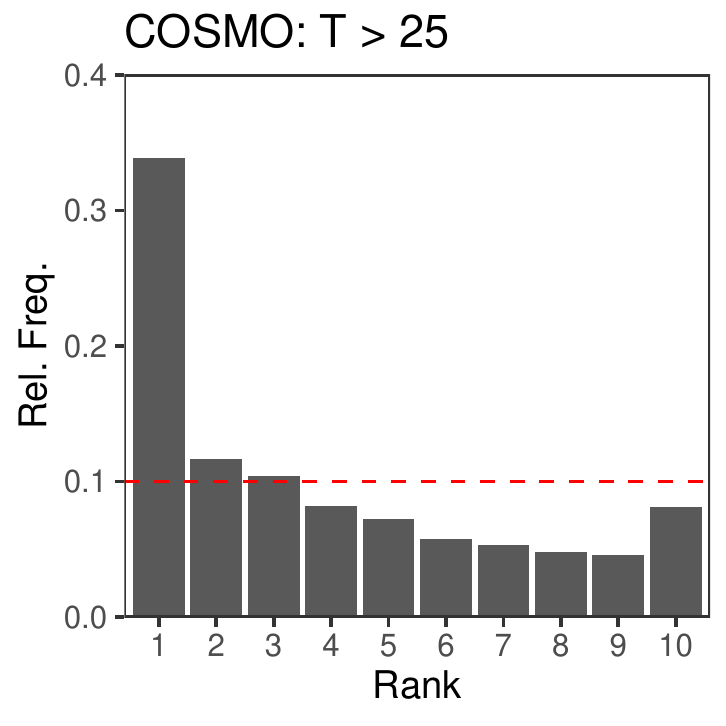}
    \raisebox{-0.5cm}{\includegraphics[width=0.36\linewidth]{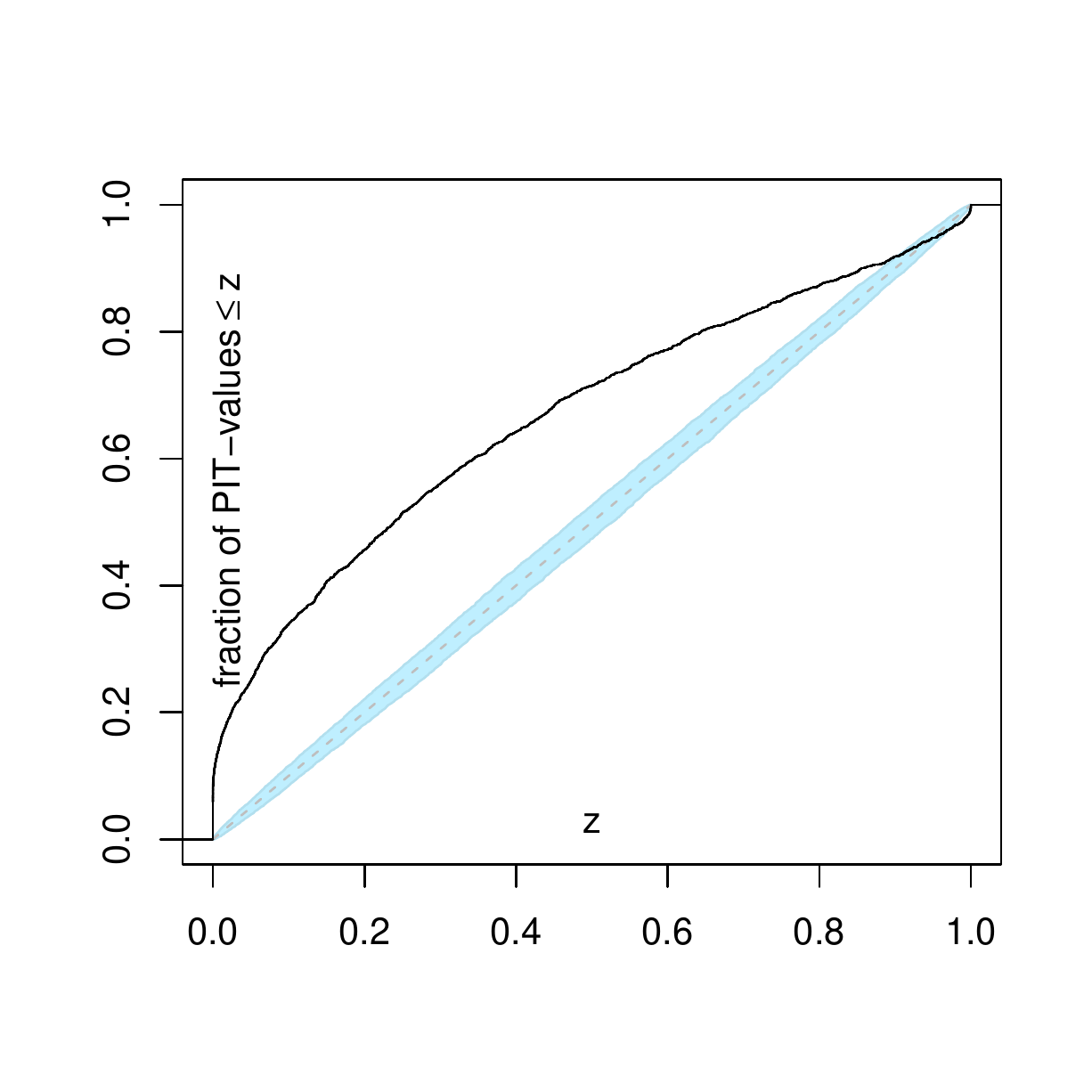}}
    \raisebox{-0.2cm}{\includegraphics[width=0.3\linewidth]{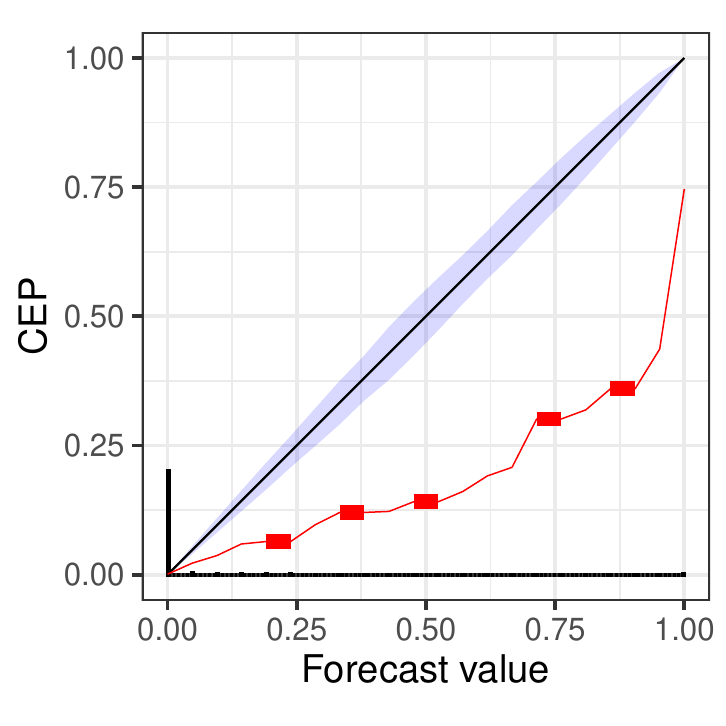}}
    \includegraphics[width=0.3\linewidth]{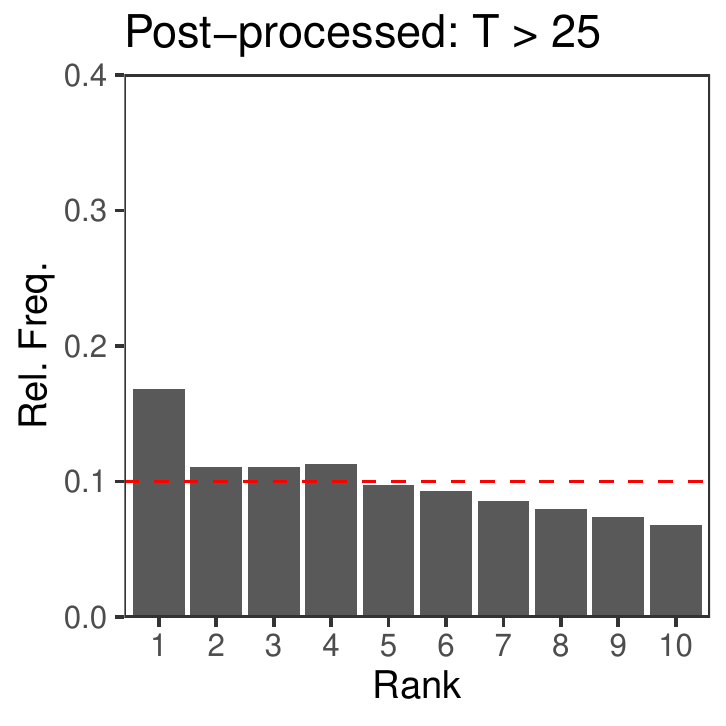}
    \raisebox{-0.5cm}{\includegraphics[width=0.36\linewidth]{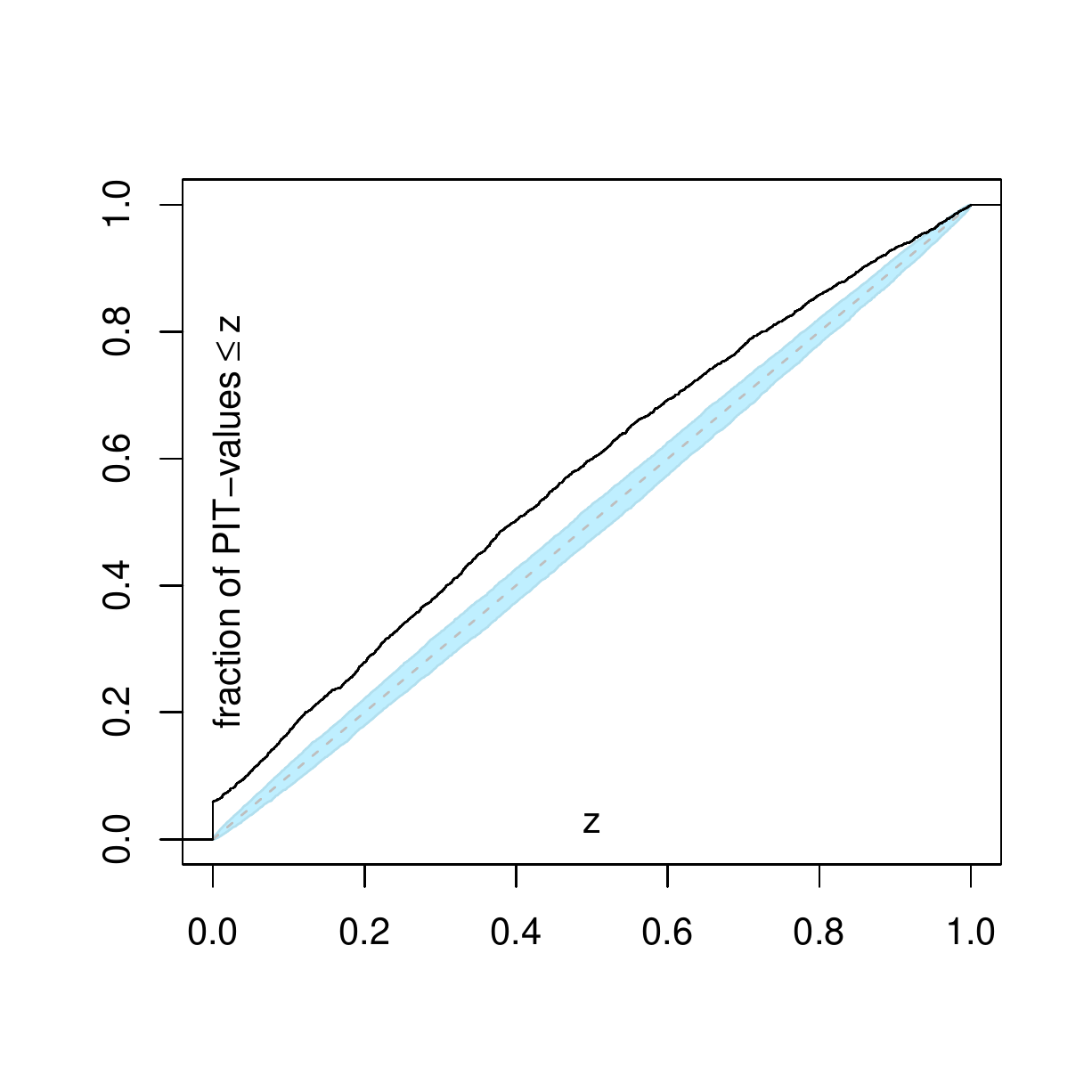}}
    \raisebox{-0.2cm}{\includegraphics[width=0.3\linewidth]{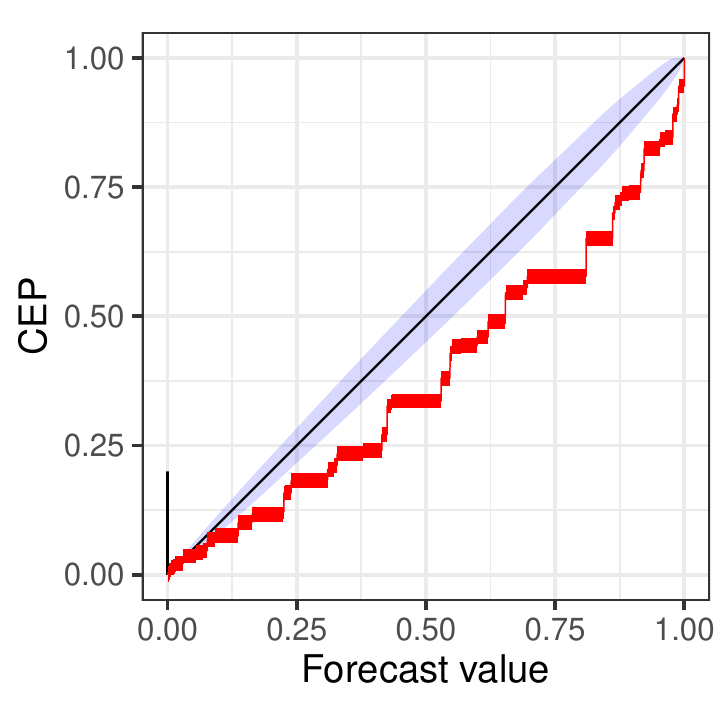}}
    \caption{Conditional PIT histograms (left) and conditional PIT reliability diagrams (middle) for the three forecasting strategies at a lead time of three days. Emphasis is on daily mean temperatures that exceed $25^{\circ}$C. Standard reliability diagrams (right) also show the conditional event probabilities (CEP) given the forecast probability that the threshold will be exceeded.}
    \label{fig:WeightedRankHists25}
\end{figure}

\begin{figure}[!t]
    \centering
    \includegraphics[width=0.3\linewidth]{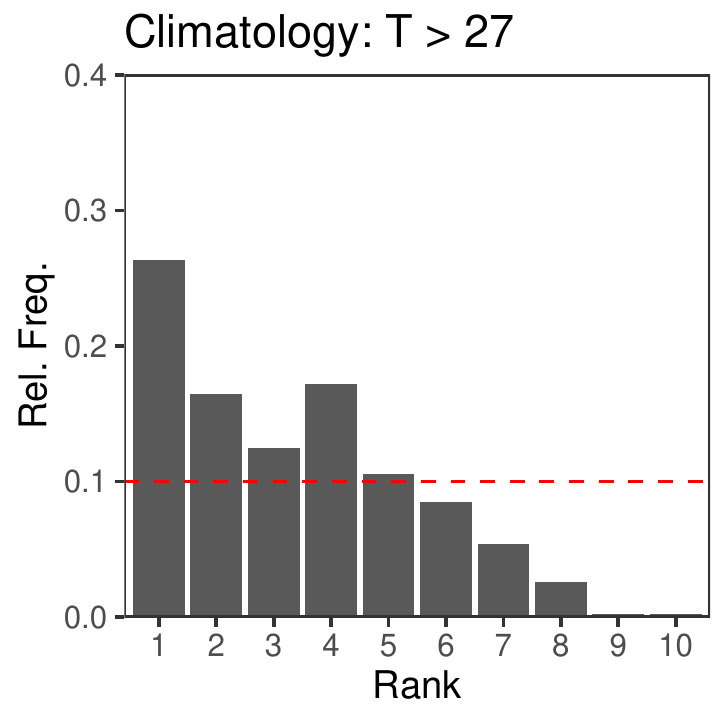}
    \raisebox{-0.5cm}{\includegraphics[width=0.36\linewidth]{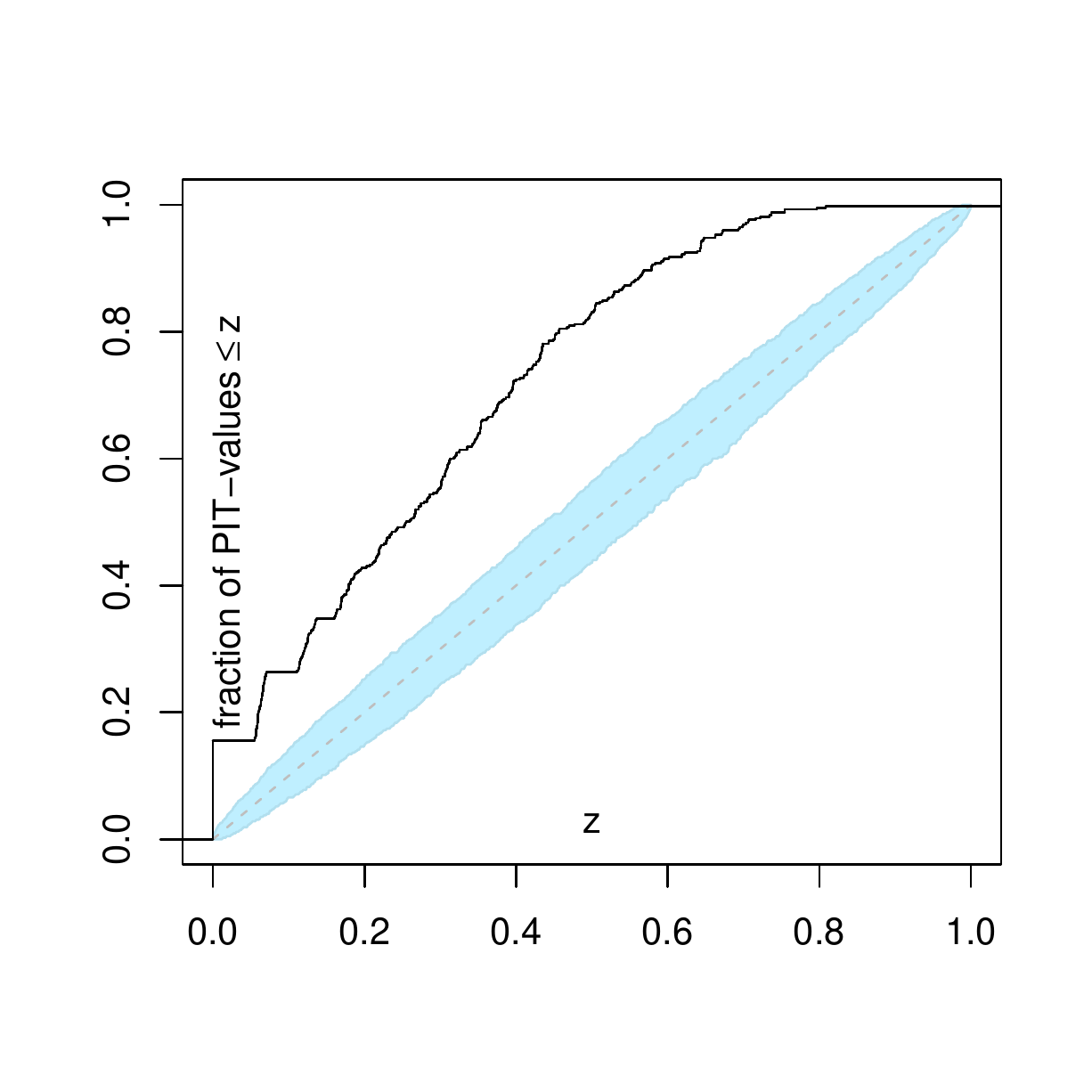}}
    \raisebox{-0.2cm}{\includegraphics[width=0.3\linewidth]{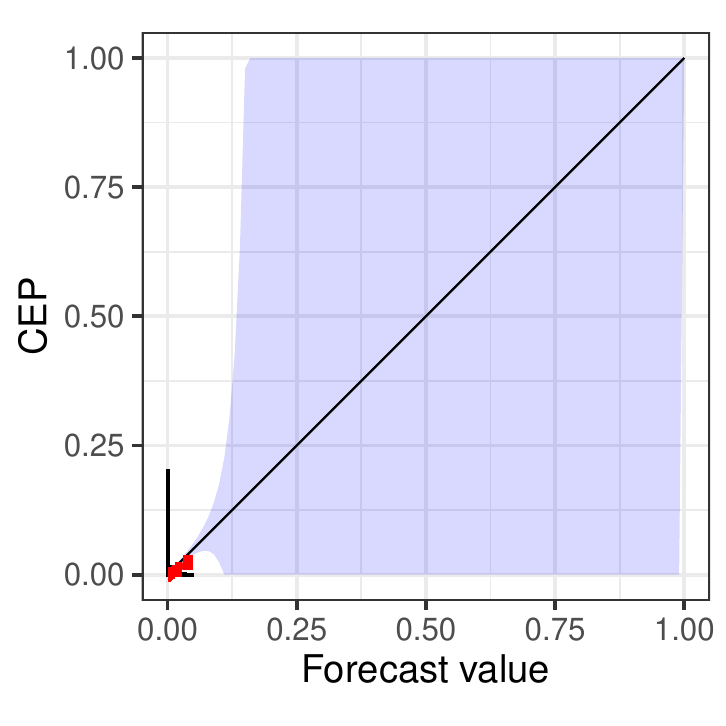}}
    \includegraphics[width=0.3\linewidth]{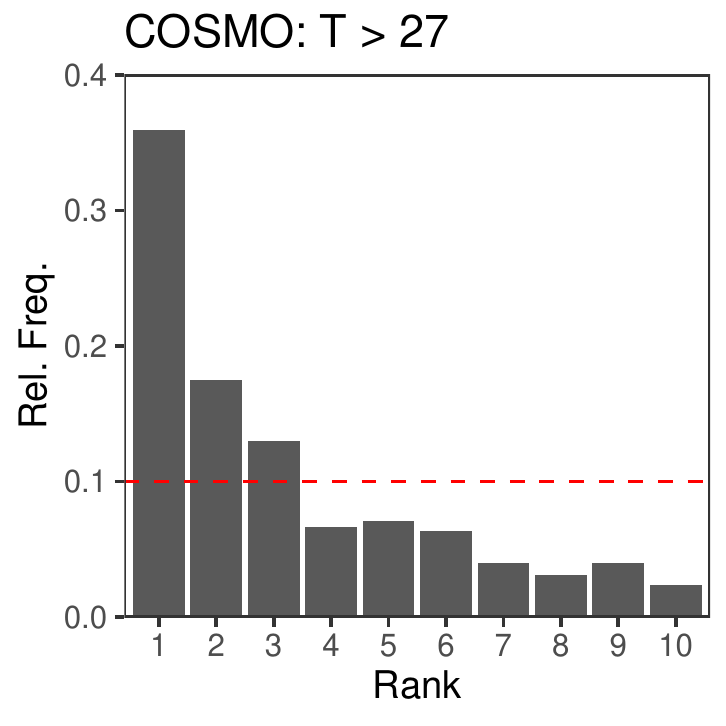}
    \raisebox{-0.5cm}{\includegraphics[width=0.36\linewidth]{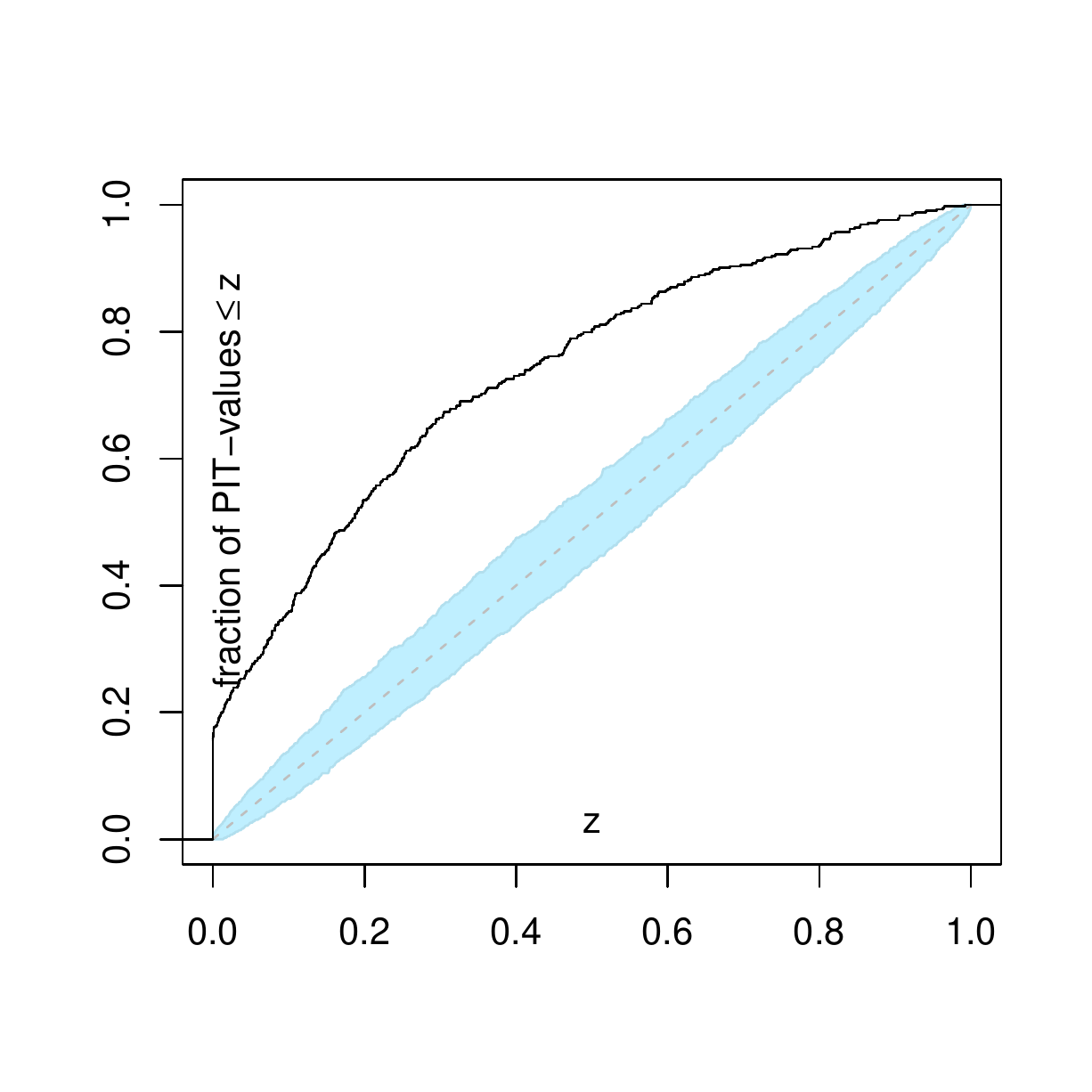}}
    \raisebox{-0.2cm}{\includegraphics[width=0.3\linewidth]{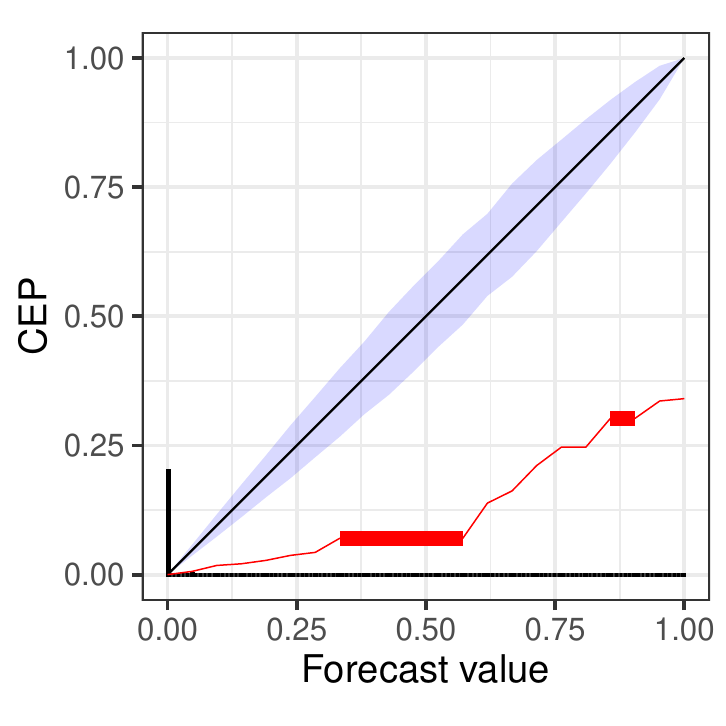}}
    \includegraphics[width=0.3\linewidth]{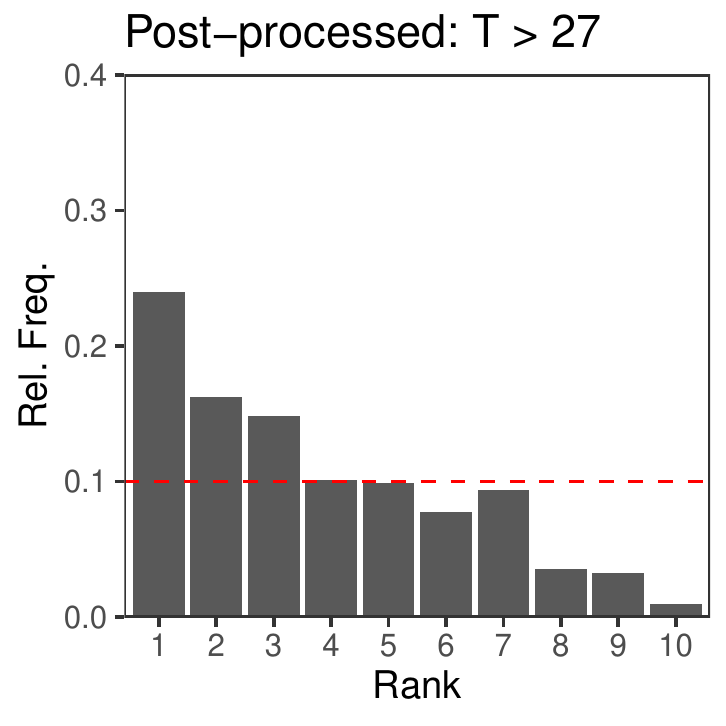}
    \raisebox{-0.5cm}{\includegraphics[width=0.36\linewidth]{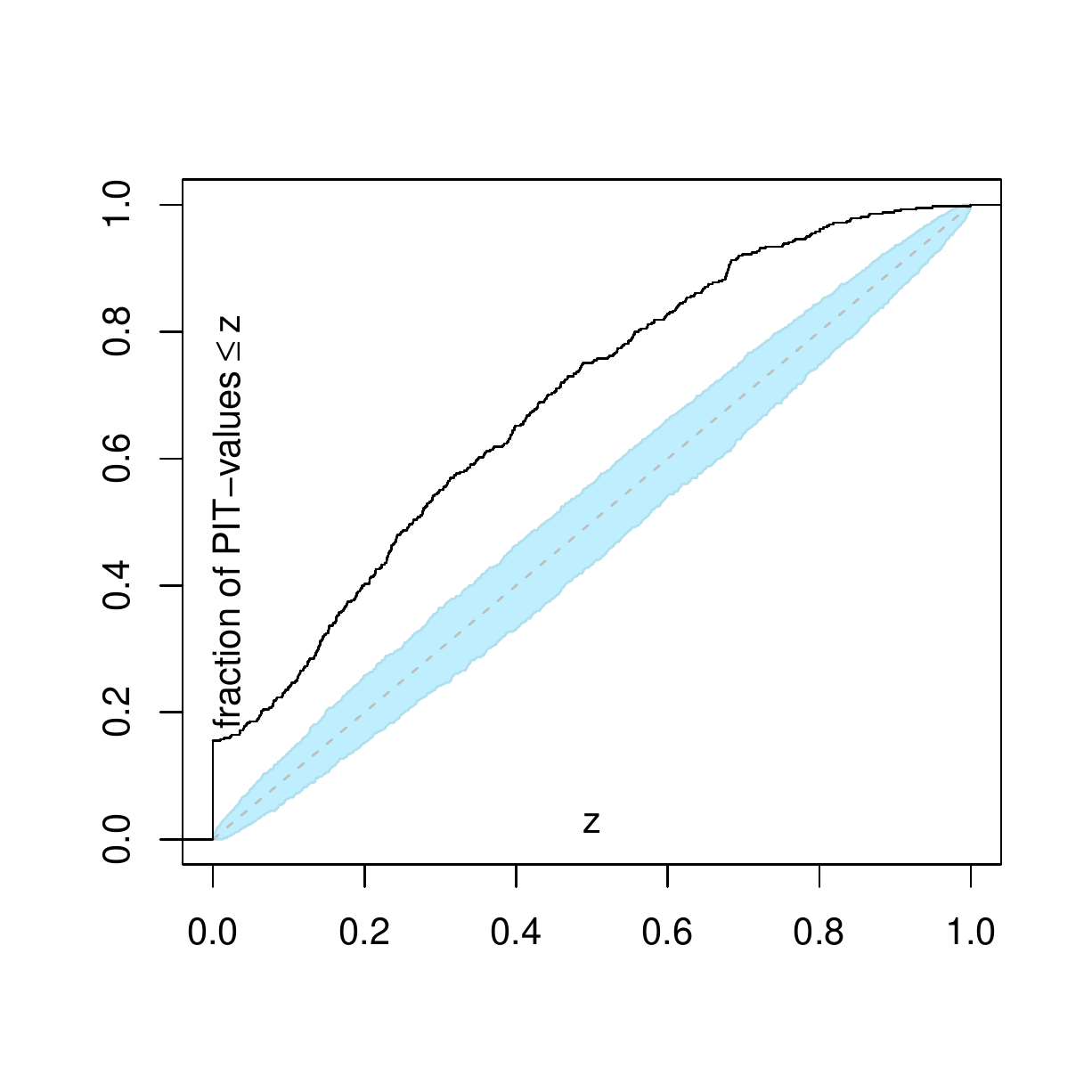}}
    \raisebox{-0.2cm}{\includegraphics[width=0.3\linewidth]{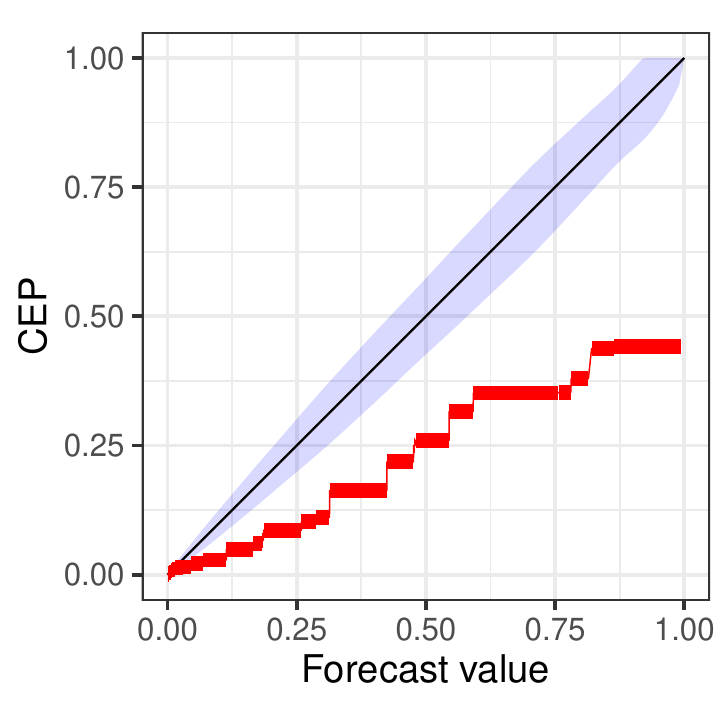}}
    \caption{As in Figure \ref{fig:WeightedRankHists25} but with emphasis on daily mean temperatures that exceed $27^{\circ}$C.}
    \label{fig:WeightedRankHists27}
\end{figure}

However, for all weighted versions of the CRPS, the skill score increases as higher thresholds are considered, suggesting the COSMO-E forecasts perform particularly poorly when predicting these more extreme events. This is true not only for the post-processed forecasts, but also for the climatological predictions. The COSMO-E forecasts, and hence also the post-processed forecasts to a lesser degree, tend to over-predict exceedances of extreme thresholds (see Figures \ref{fig:WeightedRankHists25} and \ref{fig:WeightedRankHists27}). However, the extreme thresholds are rarely exceeded by the observations, meaning the average weighted scores for the climatological forecasts are very close to zero. As a result, the climatological forecasts improve even upon the post-processed forecasts at very extreme temperature thresholds, with skill scores that tend towards one.

\bigskip

When evaluating the three competing prediction systems with respect to multivariate high-impact events, separate weight functions are chosen to emphasise the different heat levels. For example, when interest is on level two heat events, the weight is equal to one when the level two criteria in Table \ref{tab:WarningLevels} are satisfied, and zero otherwise. Equation \ref{equation:chaining} is then used to construct a chaining function for the threshold-weighted energy and variogram scores from this weight function, with $ z_{0} = (25, 25, 25) $ for heat levels one, two, and three, and $ z_{0} = (27, 27, 27) $ for heat level four.

\bigskip

For concision, only the threshold-weighted scores are presented here. The outcome-weighted scores cannot be readily applied to the multivariate ensemble forecasts without some appropriate smoothing, as discussed previously, while the vertically re-scaled scores are equivalent to the threshold-weighted scores for appropriate choices of $ x_{0} $. Of course, the weighted scores could be calculated using alternative weight functions, though in this example there are fixed definitions of extreme heat events, providing obvious weight functions with which to emphasise these events when calculating multivariate forecast accuracy. 

\bigskip

The threshold-weighted energy and variogram scores with the above weight and chaining functions are displayed in Table \ref{tab:wScores}. The scores corresponding to level one heat events are similar to those obtained from the unweighted ES and VS, while the climatological forecasts appear to perform best with respect to the most extreme heat level. COSMO-E forecasts tend to be considerably less accurate than the alternative strategies when predicting the severity of level three or four heat events.

\bigskip

\begin{table}[!b]
    \centering
    \begin{tabular}{| c | c c c c | c c c c |} 
        \hline
          & \multicolumn{4}{| c |}{twES} & \multicolumn{4}{| c |}{twVS} \\
          \cline{2-9}
          Level: & 1 & 2 & 3 & 4 & 1 & 2 & 3 & 4 \\
          \hline
          Clim. & 4.65 & 0.92 & 0.12 & 0.07 & 4.66 & 3.96 & 0.25 & 0.16 \\
          COSMO & 1.99 & 0.87 & 0.16 & 0.41 & 1.53 & 3.25 & 0.35 & 0.78\\
          Post-proc. & 1.75 & 0.71 & 0.10 & 0.11 & 1.47 & 2.55 & 0.21 & 0.23 \\
         \hline
    \end{tabular}
    \caption{Threshold-weighted ES and VS for the three forecasting strategies with emphasis on each heat event level. The weight and chaining functions used within the scores are discussed in the text. For readibility, all scores for level two and three heat events have been scaled by 10, and those for level four by 100.}
    \label{tab:wScores}
\end{table}

Conditional PIT histograms and conditional PIT reliability diagrams for the post-processed forecasts are displayed in Figures \ref{fig:WeightedRankHists25} and \ref{fig:WeightedRankHists27}, where interest is on instances where the temperature exceeds $25^{\circ}$C or $27^{\circ}$C, respectively. These checks for conditional calibration are accompanied by standard reliability diagrams for predictions that these thresholds will be exceeded. As with the owCRPS, the COSMO-E ensembles are first smoothed using a normal distribution. Figures \ref{fig:WeightedRankHists25} and \ref{fig:WeightedRankHists27} suggest that the COSMO-E ensembles over-estimate both the occurrence and severity of high temperature events, particularly for the more extreme threshold. This behaviour is also observed for the post-processed forecasts, albeit to a lesser degree. 

\bigskip

The climatological forecasts appear to issue better-calibrated forecasts for the probability of an extreme temperature event occurring, though the range of the predictions issued is much smaller than the COSMO-E and post-processed forecasts, highlighting that the climatological forecasts are less discriminative. The climatological forecast distributions also exhibit a heavy tail, suggesting parametric families other than the normal distribution may be more appropriate when modelling summer-time temperatures \citep{Allen2021b}.

\bigskip

\section{Conclusions}
\label{section:conclusion}

If meteorological services could accurately and reliably predict extreme weather events, then the impacts associated with these events could be mitigated through the design of effective warning systems. Methods to evaluate forecasts made for extreme weather are therefore crucial when developing warning systems. This paper has reviewed techniques to evaluate forecasts for high-impact events, highlighting in particular how weighted verification tools allow certain outcomes to be emphasised during forecast evaluation. We review and compare approaches to construct weighted scoring rules, both in a univariate and multivariate setting, and we then leverage the existing theory on weighted scoring rules to introduce diagnostic checks that assess forecast calibration conditionally on particular outcomes having occurred. To illustrate how these verification tools can be employed in practice, they are used to assess how well operational weather forecasts can predict dangerous heat events, defined using criteria adopted by the Swiss Federal Office of Meteorology and Climatology (MeteoSwiss). 

\bigskip

Three alternative methods to construct weighted scoring rules are compared: threshold-weighted, outcome-weighted, and vertically re-scaled scores. We argue that outcome-weighted scores provide a direct and intuitive way to circumvent the forecaster's dilemma, allowing forecasts to be evaluated only when high-impact events occur. However, when forecasts are in the form of an ensemble, and interest is on rare events, these scores are not always well-defined, making it difficult to implement them in practice. Hence, when interest is on high-impact weather events, we instead recommend evaluating forecast accuracy using threshold-weighted and vertically re-scaled scoring rules.

\bigskip

In Section \ref{section:casestudy}, we use these weighted scoring rules to evaluate three competing prediction systems whilst emphasising extreme heat events in Switzerland. In particular, we compare an operational, high-resolution ensemble prediction system to climatological and statistically post-processed forecasts. Although recent studies have suggested that statistical post-processing methods could deteriorate the accuracy of forecasts issued by numerical weather models when interest is on high-impact weather events, our results indicate that even simple post-processing methods can significantly improve upon the raw model output when predicting extreme heat in Switzerland. This suggests that forecasters should utilise statistically post-processed forecasts when constructing weather warnings, in addition to the raw output from numerical weather models. 

\bigskip

However, the checks for conditional calibration introduced here - namely conditional PIT histograms and conditional PIT reliability diagrams - indicated that even the post-processed forecasts were not calibrated conditionally on extreme temperatures having occurred, despite the overall forecast distributions being reasonably well-calibrated. Future work might therefore look to remedy this, by considering how statistical post-processing methods can be developed that are tailored to the generation of weather warnings.

\bigskip

The conditional calibration of the three prediction systems was only evaluated in the univariate setting. In Section \ref{section:verification}\ref{section:calibration}, we also describe how multivariate cPIT histograms and cPIT reliability diagrams could be constructed to check for multivariate calibration given that a high-impact event has occurred. However, as with outcome-weighted scoring rules, the approach has practical limitations when interest is on rare events and the forecast is an ensemble, which is frequently the case for multivariate weather forecasts. It would therefore be useful to design appropriate methods to convert multivariate ensemble forecasts to continuous forecast distributions, thereby allowing multivariate cPIT histograms and reliability diagrams to be applied.

\bigskip

Lastly, we reiterate that the methods discussed herein do not evaluate warning systems, but rather the ability of weather forecasts to predict potentially impactful events. Weather warnings rely not only on these forecasts, but also on several other factors: for example, the economic costs associated with a warning, the expected behaviour in response to the warning, and the effectiveness with which the warnings are relayed to those at risk. Although this renders the evaluation of weather warnings a multifaceted and thus complex task, methods to objectively identify effective warning systems would be highly valuable to operational forecasters. Future work might therefore look at developing methods to evaluate the quality of weather warnings, potentially building on the approaches presented herein to do so.

\bigskip

\section*{Acknowledgements}
This work was funded by the Swiss Federal Office of Meteorology and Climatology (MeteoSwiss) and the Oeschger Centre for Climate Change Research. We are also grateful for the helpful input provided by David Ginsbourger, Pascal Horton, Lionel Moret, Mark Liniger and Jos{\'e} Carlos Araujo Acu{\~n}a.

\section*{Data availability statement}
The code used in this study is available on GitHub at \url{https://github.com/sallen12}. The temperature observations used herein are publicly available for research purposes from MeteoSwiss' IDAweb platform, while the corresponding COSMO-E forecasts are stored in MeteoSwiss' forecast archive. Note, however, that the forecast data belongs to MeteoSwiss and is therefore not freely available. 

\bibliographystyle{apalike}
\bibliography{references}

\begin{thebibliography}{}

\bibitem[Allen et~al., 2021a]{Allen2021b}
Allen, S., Evans, G.~R., Buchanan, P., and Kwasniok, F. (2021a).
\newblock Accounting for skew when postprocessing {MOGREPS-UK} temperature
  forecast fields.
\newblock {\em Monthly Weather Review}, 149:2835--2852.

\bibitem[Allen et~al., 2021b]{Allen2021}
Allen, S., Evans, G.~R., Buchanan, P., and Kwasniok, F. (2021b).
\newblock Incorporating the {North Atlantic Oscillation} into the
  post-processing of {MOGREPS-G} wind speed forecasts.
\newblock {\em Quarterly Journal of the Royal Meteorological Society},
  147:1403--1418.

\bibitem[Allen et~al., 2022]{Allen2022}
Allen, S., Ginsbourger, D., and Ziegel, J. (2022).
\newblock Evaluating forecasts for high-impact events using transformed kernel
  scores.
\newblock {\em arXiv preprint arXiv:2202.12732}.

\bibitem[Arnold et~al., 2021]{Arnold2021}
Arnold, S., Henzi, A., and Ziegel, J.~F. (2021).
\newblock Sequentially valid tests for forecast calibration.
\newblock {\em arXiv preprint arXiv:2109.11761}.

\bibitem[Basaga{\~n}a et~al., 2011]{Basagana2011}
Basaga{\~n}a, X., Sartini, C., Barrera-G{\'o}mez, J., Dadvand, P., Cunillera,
  J., Ostro, B., Sunyer, J., and Medina-Ram{\'o}n, M. (2011).
\newblock Heat waves and cause-specific mortality at all ages.
\newblock {\em Epidemiology}, 22:765--772.

\bibitem[Bellier et~al., 2017]{Bellier2017}
Bellier, J., Zin, I., and Bontron, G. (2017).
\newblock Sample stratification in verification of ensemble forecasts of
  continuous scalar variables: Potential benefits and pitfalls.
\newblock {\em Monthly Weather Review}, 145:3529--3544.

\bibitem[Brier, 1950]{Brier1950}
Brier, G.~W. (1950).
\newblock Verification of forecasts expressed in terms of probability.
\newblock {\em Monthly Weather Review}, 78:1--3.

\bibitem[Coles et~al., 2001]{Coles2001}
Coles, S., Bawa, J., Trenner, L., and Dorazio, P. (2001).
\newblock {\em An introduction to statistical modeling of extreme values},
  volume 208.
\newblock London: Springer.

\bibitem[Dawid, 1984]{Dawid1984}
Dawid, A.~P. (1984).
\newblock Statistical theory: {The} prequential approach.
\newblock {\em Journal of the Royal Statistical Society: Series A (General)},
  147:278--290.

\bibitem[Dawid and Sebastiani, 1999]{DawidSebastiani1999}
Dawid, A.~P. and Sebastiani, P. (1999).
\newblock Coherent dispersion criteria for optimal experimental design.
\newblock {\em Annals of Statistics}, pages 65--81.

\bibitem[Delle~Monache et~al., 2006]{DelleMonache2006}
Delle~Monache, L., Hacker, J.~P., Zhou, Y., Deng, X., and Stull, R.~B. (2006).
\newblock Probabilistic aspects of meteorological and ozone regional ensemble
  forecasts.
\newblock {\em Journal of Geophysical Research: Atmospheres}, 111.

\bibitem[Diks et~al., 2011]{Diks2011}
Diks, C., Panchenko, V., and Van~Dijk, D. (2011).
\newblock Likelihood-based scoring rules for comparing density forecasts in
  tails.
\newblock {\em Journal of Econometrics}, 163:215--230.

\bibitem[Dimitriadis et~al., 2021]{Dimitriadis2021}
Dimitriadis, T., Gneiting, T., and Jordan, A.~I. (2021).
\newblock Stable reliability diagrams for probabilistic classifiers.
\newblock {\em Proceedings of the National Academy of Sciences}, 118.

\bibitem[Ferro and Stephenson, 2011]{Ferro2011}
Ferro, C. A.~T. and Stephenson, D.~B. (2011).
\newblock Extremal dependence indices: Improved verification measures for
  deterministic forecasts of rare binary events.
\newblock {\em Weather and Forecasting}, 26:699--713.

\bibitem[Gneiting et~al., 2007]{Gneiting2007}
Gneiting, T., Balabdaoui, F., and Raftery, A.~E. (2007).
\newblock Probabilistic forecasts, calibration and sharpness.
\newblock {\em Journal of the Royal Statistical Society: Series B (Statistical
  Methodology)}, 69:243--268.

\bibitem[Gneiting and Raftery, 2007]{GneitingRaftery2007}
Gneiting, T. and Raftery, A.~E. (2007).
\newblock Strictly proper scoring rules, prediction, and estimation.
\newblock {\em Journal of the American Statistical Association}, 102:359--378.

\bibitem[Gneiting et~al., 2005]{Gneiting2005}
Gneiting, T., Raftery, A.~E., Westveld~III, A.~H., and Goldman, T. (2005).
\newblock Calibrated probabilistic forecasting using ensemble model output
  statistics and minimum {CRPS} estimation.
\newblock {\em Monthly Weather Review}, 133:1098--1118.

\bibitem[Gneiting and Ranjan, 2011]{GneitingRanjan2011}
Gneiting, T. and Ranjan, R. (2011).
\newblock Comparing density forecasts using threshold-and quantile-weighted
  scoring rules.
\newblock {\em Journal of Business \& Economic Statistics}, 29:411--422.

\bibitem[Gneiting and Resin, 2021]{Gneiting2022}
Gneiting, T. and Resin, J. (2021).
\newblock Regression diagnostics meets forecast evaluation: Conditional
  calibration, reliability diagrams, and coefficient of determination.
\newblock {\em arXiv preprint arXiv:2108.03210}.

\bibitem[Hamill, 2001]{Hamill2001}
Hamill, T.~M. (2001).
\newblock Interpretation of rank histograms for verifying ensemble forecasts.
\newblock {\em Monthly Weather Review}, 129:550--560.

\bibitem[Hamill and Colucci, 1997]{Hamill1997}
Hamill, T.~M. and Colucci, S.~J. (1997).
\newblock Verification of {Eta--RSM} short-range ensemble forecasts.
\newblock {\em Monthly Weather Review}, 125:1312--1327.

\bibitem[Holzmann and Klar, 2017]{Holzmann2017}
Holzmann, H. and Klar, B. (2017).
\newblock Focusing on regions of interest in forecast evaluation.
\newblock {\em The Annals of Applied Statistics}, 11:2404--2431.

\bibitem[Jolliffe and Stephenson, 2012]{Jolliffe2012}
Jolliffe, I.~T. and Stephenson, D.~B. (2012).
\newblock {\em Forecast verification: a practitioner's guide in atmospheric
  science}.
\newblock John Wiley \& Sons.

\bibitem[Keller et~al., 2021]{Keller2021}
Keller, R., Rajczak, J., Bhend, J., Spirig, C., Hemri, S., Liniger, M.~A., and
  Wernli, H. (2021).
\newblock Seamless multimodel postprocessing for air temperature forecasts in
  complex topography.
\newblock {\em Weather and Forecasting}, 36:1031--1042.

\bibitem[Lerch and Thorarinsdottir, 2013]{Lerch2013}
Lerch, S. and Thorarinsdottir, T.~L. (2013).
\newblock Comparison of non-homogeneous regression models for probabilistic
  wind speed forecasting.
\newblock {\em Tellus A}, 65:21206.

\bibitem[Lerch et~al., 2017]{Lerch2017}
Lerch, S., Thorarinsdottir, T.~L., Ravazzolo, F., and Gneiting, T. (2017).
\newblock Forecaster's dilemma: {Extreme} events and forecast evaluation.
\newblock {\em Statistical Science}, 32:106--127.

\bibitem[Matheson and Winkler, 1976]{Matheson1976}
Matheson, J.~E. and Winkler, R.~L. (1976).
\newblock Scoring rules for continuous probability distributions.
\newblock {\em Management Science}, 22:1087--1096.

\bibitem[Pantillon et~al., 2018]{Pantillon2018}
Pantillon, F., Lerch, S., Knippertz, P., and Corsmeier, U. (2018).
\newblock Forecasting wind gusts in winter storms using a calibrated
  convection-permitting ensemble.
\newblock {\em Quarterly Journal of the Royal Meteorological Society},
  144:1864--1881.

\bibitem[Pinson and Tastu, 2013]{Pinson2013}
Pinson, P. and Tastu, J. (2013).
\newblock Discrimination ability of the energy score.
\newblock Technical report, Technical University of Denmark.

\bibitem[Ragettli et~al., 2017]{Ragettli2017}
Ragettli, M.~S., Vicedo-Cabrera, A.~M., Schindler, C., and R{\"o}{\"o}sli, M.
  (2017).
\newblock Exploring the association between heat and mortality in {Switzerland}
  between 1995 and 2013.
\newblock {\em Environmental Research}, 158:703--709.

\bibitem[Schefzik et~al., 2013]{Schefzik2013}
Schefzik, R., Thorarinsdottir, T.~L., and Gneiting, T. (2013).
\newblock Uncertainty quantification in complex simulation models using
  ensemble copula coupling.
\newblock {\em Statistical Science}, 28:616--640.

\bibitem[Scheuerer and Hamill, 2015]{ScheuererHamill2015}
Scheuerer, M. and Hamill, T.~M. (2015).
\newblock Variogram-based proper scoring rules for probabilistic forecasts of
  multivariate quantities.
\newblock {\em Monthly Weather Review}, 143:1321--1334.

\bibitem[Stephenson et~al., 2008]{Stephenson2008}
Stephenson, D.~B., Casati, B., Ferro, C., and Wilson, C. (2008).
\newblock The extreme dependency score: A non-vanishing measure for forecasts
  of rare events.
\newblock {\em Meteorological Applications}, 15:41--50.

\bibitem[Thorarinsdottir and Schuhen, 2018]{Thorarinsdottir2018book}
Thorarinsdottir, T.~L. and Schuhen, N. (2018).
\newblock Verification: Assessment of calibration and accuracy.
\newblock In {\em Statistical postprocessing of ensemble forecasts}, pages
  155--186. Elsevier.

\bibitem[Vannitsem et~al., 2018]{Vannitsem2018}
Vannitsem, S., Wilks, D.~S., and Messner, J. (2018).
\newblock {\em Statistical postprocessing of ensemble forecasts}.
\newblock Elsevier.

\bibitem[Wilks, 2019]{Wilks2019}
Wilks, D.~S. (2019).
\newblock {\em Statistical methods in the atmospheric sciences}.
\newblock Amsterdam: Elsevier.

\bibitem[WMO, 2015]{WMO2015}
WMO (2015).
\newblock {WMO} guidelines on multi-hazard impact-based forecast and warning
  services.

\bibitem[Ziegel, 2017]{Ziegel2017}
Ziegel, J. (2017).
\newblock Copula calibration.
\newblock In {\em Copulae: On the Crossroads of Mathematics and Economics},
  pages 7--10. Mathematisches Forschungsinstitut Oberwolfach. Report No.
  20/2015.

\bibitem[Zscheischler et~al., 2020]{Zscheischler2020}
Zscheischler, J., Martius, O., Westra, S., Bevacqua, E., Raymond, C., Horton,
  R.~M., van~den Hurk, B., AghaKouchak, A., J{\'e}z{\'e}quel, A., Mahecha,
  M.~D., Maraun, D., Ramos, A.~M., Ridder, N.~N., Thiery, W., and Vignotto, E.
  (2020).
\newblock A typology of compound weather and climate events.
\newblock {\em Nature Reviews Earth \& Environment}, 1:333--347.

\end{thebibliography}

\end{document}